**A Probabilistic Case For A Large Missing Carbon Sink On Mars After 3.5 Billion Years Ago**

Andy W. Heard[1,2,*], and Edwin S. Kite[1]

1. Department of the Geophysical Sciences, The University of Chicago, 5734 South Ellis Avenue, Chicago, IL 60637, United States.

2. Origins Laboratory, Department of the Geophysical Sciences and Enrico Fermi Institute, The University of Chicago, 5734 South Ellis Avenue, Chicago, IL 60637, United States.

*Corresponding author. Email: andyheard@uchicago.edu

Highlights

- We combine O loss, H isotopes and soil reservoirs to estimate Martian paleo-$pCO_2$ at 3.5 Ga
- Soil reservoirs are a small O sink relative to escape to space
- Combined O sinks for post-3.5 Ga Mars are inconsistent with high paleo-$pCO_2$
- There is strong likelihood that there is a missing C sink on Mars after 3.5 Ga

Abstract word count: 266

Main text word count: 6396

**Abstract**

Mars has a thin (6 mbar) $CO_2$ atmosphere currently. There is strong evidence for paleolakes and rivers formed by warm climates on Mars, including after 3.5 billion years (Ga) ago, which indicates that a $CO_2$ atmosphere thick enough to permit a warm climate was present at these times. Since Mars no longer has a thick $CO_2$ atmosphere, it must have been lost. One possibility is that Martian $CO_2$ was lost to space. Oxygen escape rates from Mars are high enough to account for loss of a thick $CO_2$ atmosphere, if $CO_2$ was the main source of escaping O. But here, using H isotope ratios, O escape calculations, and quantification of the surface O sinks on Mars, we show for the first time that O escape from Mars after 3.5 Ga must have been predominantly associated with the loss of $H_2O$, not $CO_2$, and therefore it is unlikely that ≥250 mbar Martian $CO_2$ has been lost to space in the last 3.5 Ga, because such results require highly unfavored O loss scenarios. It is possible that the presence of young rivers and lakes on Mars could be reconciled with limited $CO_2$ loss to space if crater chronologies on Mars are sufficiently incorrect that all apparently young rivers and lakes are actually older than 3.5 Ga, or if climate solutions exist for sustained runoff on Mars with atmospheric $CO_2$ pressure <250 mbar. However, our preferred solution to reconcile the presence of <3.5 Gya rivers and lakes on Mars with the limited potential for $CO_2$ loss to space is a large, as yet undiscovered, geological C sink on Mars.

## 1. Introduction

Changes in Mars' carbon dioxide ($CO_2$) and water ($H_2O$) inventories are key unknown in understanding the planet's climate evolution, and constraining these changes requires knowledge of Martian historical volatile sinks (Figure 1; Catling and Kasting, 2017; Jakosky and Phillips, 2001). The Late Hesperian-Amazonian (Kite et al., 2019; see Table 1 for a summary of Martian epoch dates) poses a particular challenge for Mars' $CO_2$ evolution. Post-3.5 Ga ago river channels and lakes indicate significant flowing water, and thus sufficient $CO_2$ to permit a climate warm enough for water to flow (e.g. Dickson et al., 2009; Grant and Wilson, 2012; Irwin III et al., 2015; Kite, 2019; Kite et al., 2019; Palucis et al., 2016); but the currently postulated ≥0.25 bar $CO_2$ atmospheres required sustained warm climates (Haberle et al., 2017; Kite, 2019; Mansfield et al., 2018; Ramirez et al., 2014; Urata and Toon, 2013; Wordsworth, 2016), greatly exceed the current exchangeable inventory of 12 mbar (6 mbar in atmosphere + 6 mbar in ice caps; Putzig et al., 2018) $CO_2$. A lot of $CO_2$ must have been lost from Mars's atmosphere somehow.

The possible $CO_2$ sinks for Mars are gradual escape to space, either as $CO_2$, $CO_2^+$, or the constituent C and O atoms; or fixing of C in the Martian (sub)surface as carbonates, clathrates, or ices. After 3.5 Ga, impact erosion of the atmosphere should no longer be significant. Some of these sinks for $CO_2$ on late Hesperian-Amazonian Mars are demonstrably too small. Only a few mbar of $CO_2$ can be accounted for by 3.5 Ga-integrated $CO_2^+$ loss to space (Dong et al., 2018; Ramstad et al., 2018), and known post-Noachian carbonate sequestration was very limited in extent (Edwards and Ehlmann, 2015; Niles et al., 2013). However, the potential for loss of $CO_2$ to space as its constituent atoms is supported by significant modern-day escape of O from Mars, which has been measured by the Mars Atmosphere and Volatile EvolutioN (MAVEN) mission (Jakosky et al., 2018; Lillis et al., 2017). Furthermore, this Martian atmospheric O loss can be extrapolated to higher rates in the past using estimates of the Sun's evolving radiative flux and stellar wind, which provides energy for atmospheric escape to space. Since O is a very minor constituent of Mars' atmosphere, this O must be ultimately sourced from $CO_2$ or $H_2O$. There is degeneracy to the problem of O escape in that escaping O may ultimately derive from either $CO_2$ or $H_2O$ (Jakosky et al. 2018). Modern loss rate ratios for H:O are consistent with ≈2:1 and suggest that at present, most O being lost from Mars derives from $H_2O$ (Jakosky et al., 2018). It is highly uncertain whether this coupling has held at a constant value over geological timescales, or even whether it holds on year-to-year dust storm-influenced or 11-year Solar cycle-influenced timescales (Jakosky et al., 2018). Meanwhile, the fraction of $H_2O$- or $CO_2$-derived O that went into oxidation of soil and sediments rather than escaping to space (Lammer et al., 2003) is not well constrained. These uncertainties leave the possibility that a large portion of the historical O loss flux corresponds to $CO_2$, which would imply a major loss-to-space channel for C or CO (ions or neutrals) was active on Mars after 3.5 Ga ago.

Complementing the new constraints on O loss from Mars that MAVEN provides, new constraints on Mars's H loss history come from recent hydrogen isotope ratio (D/H) data. The Mars Science Laboratory (MSL) mission's Curiosity Rover has provided modern and ancient (≈ 3.5 Ga) datapoints for Martian D/H in the modern atmosphere (Webster et al., 2013; Villanueva et al., 2015) and ancient hydrosphere (Mahaffy et al., 2015). The ancient D/H measurement is from a lacustrine mudstone, and therefore samples the hydrosphere at a time when lakes were present (Mahaffy et al., 2015). D/H data show that the Martian H reservoir has become isotopically heavier over time, which implies the preferential loss of isotopically lighter H to space over time, with a greater change in D/H implicating more severe loss of H (Usui 2019). These observations can be used to estimate historical H loss from Mars after 3.5 Ga ago. In this way, the rover data bypasses the uncertainty in whether or not MAVEN-era H escape is representative of H escape over 11-year and longer (geological) timescales.

Finally, better maps of Mars's post-3.5 Ga sedimentary deposits (Michalski and Niles, 2012), and better estimates of the true regolith depth on Mars (Warner et al., 2017), make it easier to quantify the sedimentary O sink.

We ask if the three independent datasets; on O escape, D/H evolution, and the soil and sedimentary reservoir; reveal C escape from Mars over the last 3.5 Ga at a sufficient rate to account for the loss of a ≥250 mbar $CO_2$ atmosphere. In this work we combine constraints on O and H loss from Mars to extract the O loss corresponding to $CO_2$ (Figure 1). Then, we use an error-propagation approach to determine upper limits on 3.5 Ga $pCO_2$ allowed by plausible O-loss and hydrosphere evolution scenarios (Figure 1b). Our model includes an improved accounting for post-3.5 Ga soil and sedimentary oxidative O sinks. We describe the model approach and input parameters in Section 2. We present model results in Section 3. We discuss the model results and their implications in Section 4. Section 5 contains our conclusions. Our main results are shown in Figures 2, 3, 4 and 6.

## 2. Model Description

### 2.1. Extracting $CO_2$ loss from O and H constraints

The expression for the upper limit of $CO_2$-derived C atoms escaped from Mars to space is

$$\#C_{esc,max} = 1/2 \times \left( \#O_{esc,max} - \left( 1/2 \times \#H_{esc,min} - \#O_{ox,max} \right) \right) + \#CO_2^+{}_{esc,max} \quad (1)$$

where #C, #O, and #H, and $\#CO_2^+$ are the number of C, O and H atoms and $CO_2^+$ ions that are lost from the Martian hydrosphere-atmosphere system through escape to space, or oxidation of surficial rock reservoirs, denoted by the subscripts $_{esc}$, and $_{ox}$, respectively (Figure 1). The expression dictates that the maximum $CO_2$-derived C lost to space over a given time interval is half (due to the stoichiometry of $CO_2$) of the amount of O escaping over that time interval that cannot be attributed to escape of $H_2O$-derived

O. The loss of $H_2O$-derived O is independently constrained by the escape of H atoms determined from D/H evolution on Mars, while accounting for the fraction of O which was sequestered in Martian soil and layered sedimentary deposits by oxidation (Figure 1). The expression above does not account for sequestration of $CO_2$ in Mars's crust but such additional sinks are also considered in our discussion. We also do not consider the additional sinks for Martian water that we consider unlikely to be strongly fractionating for H isotopes. For example, crustal hydration reactions such as clay mineral formation could constitute a major sink for Martian water (Wernicke and Jakosky, 2019), but clay mineral formation would not cause a several-times enrichment in D in the residual Martian hydrosphere (e.g. Liu and Epstein, 1984). Furthermore, clay mineral formation at the scale likely to influence the Martian planetary water budget predates the 3.5 Ga start of our model and was largely restricted to the Noachian (Ehlmann et al., 2011; Ehlmann and Edwards, 2014). Our parameterization of H escape to space is done solely to identify the water-associated loss of O from Mars since 3.5 Ga and we do not intend this work to be a complete accounting of Martian hydrological evolution.

Our model assumes a unidirectional decline of Martian $pCO_2$ as a result of sink processes operating after 3.5 Ga. We do not rule out a later introduction for some of the $CO_2$ lost on Mars after 3.5 Ga; but including it in our model could only strengthen our upper limits on post-3.5 Ga $pCO_2$ on Mars. If later outgassing occurred, this $CO_2$ was subsequently lost from known exchangeable reservoirs on Mars, in addition to the $CO_2$ that was present at 3.5 Ga. Moreover, later-outgassed $CO_2$ was vulnerable to sink processes for a shorter duration and must still be drawn down to 12 mbar equivalent present-day exchangeable $CO_2$ (Putzig et al., 2018). This places a greater demand on sink processes. In the case of escape to space, we expect any loss fluxes to be attenuating towards the present.

### 2.2. Oxygen loss to space

Our new approach to inferring post-3.5 Ga ago pCO$_2$ is made timely in part by new constraints on O loss from Mars to space provided by MAVEN measurements. Loss of O to space occurs through several loss channels, separable into photochemical escape of hot O atoms following sputtering by solar wind; pickup ion loss; and dissociative recombination (Dong et al., 2018; Jakosky et al., 2018). Dissociative recombination was recently confirmed to be the dominant modern escape route for O (Jakosky et al., 2018). The approach to calculating the integrated effect of dissociative recombination over 3.5 Ga is detailed below. Our calculations of O loss to space were performed following the approach of Lillis et al. (2017) for the main dissociative recombination loss channel, and the parameterizations of Jakosky et al. (2018) for the other loss channels of pickup O$^+$ ion loss, O sputtering, and ionospheric O$^+$ outflow. The details of these calculations are provided in the Appendix, including a discussion of the solar extreme ultraviolet (EUV) evolution that drove dissociative recombination, and a discussion of the uncertainty on various input parameters for this calculation.

## 2.3. CO$_2$ ion loss

Carbon dioxide can escape Mars directly as CO$_2^+$ ions. Early results from the Mars Express mission indicated a loss rate of 8 x 10$^{22}$ CO$_2^+$ s$^{-1}$ (Barabash et al., 2007). This number has subsequently been revised upward with MAVEN data to 3.6 x 10$^{23}$ CO$_2^+$ s$^{-1}$ (Dong et al., 2018; Ramstad et al., 2018). In this study, the modelled CO$_2^+$ loss history from Dong et al. (2018) is used as published. This loss channel is included for completeness but loss since 3.5 Ga is ≈ 2-6 mbar and therefore is not a strong control on overall results (Dong et al., 2018; Ramstad et al., 2018).

## 2.4. Surface oxygen sinks

Some oxygen which initially resided in the Martian hydrosphere (H$_2$O), or atmosphere (H$_2$O, CO$_2$), now resides in chemically weathered rocks near Mars's surface. Indeed, aqueous minerals, many of them oxidized, formed in abundance before 3.5 Ga ago (Ehlmann and Edwards, 2014). However, the present study is concerned only with oxidative alteration that occurred after 3.5 Ga ago. In this epoch, the main surficial O sinks are layered sedimentary deposits, including dust deposits lying within polar and low latitude glacial deposits; and soil. The list of deposits, their estimated mass, and the sequestered oxygen they represent, is given in Supplementary Table 1.

The reactions considered are (1) the oxidation of ferrous iron ($Fe^{2+}$) in igneous materials to ferric iron ($Fe^{3+}$) in hematite, goethite, nanophase iron oxide phases etc.; and (2) oxidation of sulfur species; primarily $SO_2$ and $H_2S$ volatiles but potentially also igneous crustal sulfides and sedimentary sulfite; to sulfate ($SO_3$). The $Fe^{2+}$ oxidation reaction consumes 0.5 moles of O per mole of Fe. The sulfur oxidation reactions require 1 mole and 4 moles of O per mole of S to oxidize $SO_2$ and $H_2S$, respectively. Both Fe and S may either be oxidized directly by oxygen liberated by photolysis in the atmosphere, or via UV-promoted photo-oxidation reactions that derive the oxygen in the stable reaction products from hydrolysis of water (Hurowitz et al., 2010). Large quantities of sulfur volatiles erupted to the Martian surface in the Late Hesperian, and some of the largest late-formed layered sedimentary deposits on Mars have a high sulfur content relative to the Martian crust and appear to derive their S from volcanic outgassing (Michalski and Niles, 2012). Because volcanically erupted S-volatiles are piped directly into the oxidizing surface environment, and S has greater reducing power per mole, sulfur provides a major O sink despite being a minor constituent of the bulk Martian crust relative to Fe. Surface oxygen sinks are assessed assuming a ferric iron content of typical Martian soils (Morris et al., 2006). We adopt $SO_3$ (sulfate in rock) content of 20 wt%, following Michalski and Niles (2012), which they assess to be a representative composition of layered sedimentary deposits, which volumetrically dominate over the global soil layer. We assume that the initial speciation of sulfur volatiles now residing in sulfate was 50% $SO_2$, 50% $H_2S$, consistent with the more reducing redox state of the Martian mantle relative to Earth's (e.g. Halevy et al., 2007). We calculate that the total soil and layered sedimentary oxidative O sink is ≈ 4.3 x $10^{19}$ moles of O, which is equivalent to a 242 mbar $pCO_2$ drawdown (Table 2; Supplementary Table 1).

*2.5. D/H ratio constraint on post 3.5 Ga water loss*

Hydrogen isotope (D/H) ratios track Martian hydrosphere evolution independent of constraints from O loss (Jakosky, 1991; Mahaffy et al., 2015). Hydrogen is preferentially lost relative to heavier deuterium during atmospheric escape. Therefore, progressive Martian D/H enrichment over time can track loss of the Martian hydrosphere. The relative size of exchangeable (surface water and exchangeable ground ice) water reservoirs at two points in time ($R_{t1}$, $R_{t2}$) relate to the D/H ratio through time ($I_{t2}$, $I_{t2}$) through the expression

$$\frac{R_{t1}}{R_{t2}} = \left(\frac{I_{t2}}{I_{t1}}\right)^{\frac{1}{1-f}} \quad (2)$$

Where *f* is the fractionation factor, and *t2* refers to a time after *t1* (Kurokawa et al., 2014). In this study, *t1* refers to 3.5 Ga, and *t2* refers to present day. The values for parameters used in this model are listed in Table 2. Accurate assessment of Martian water loss ($R_{t2} - R_{t1}$) requires constraints on the modern water reservoir, modern and ancient D/H ratios, and a reasonable value for *f*, defined by the expression

$$f = \frac{\Delta[D]/[D]}{\Delta[H]/[H]} \quad (3)$$

Where [D] and [H] are the atomic abundances of D and H respectively in the hydrosphere (Kurokawa et al., 2014). A smaller value for *f* gives greater preferential loss of H relative to D. The net fractionation factor for H escape *f* is the product ($f = f_a \times f_{esc}$) of fractionation between $H_2O$ and $H_2$, with $f_a$ = 0.41; and the effect of different escape rates for H and D, $f_{esc}$. Modern Mars has *f* = 0.016, which gives the most strongly fractionating regime for hydrogen escape that we expect; however, this corresponds to nonthermal escape processes not thought to dominate on early Mars (Krasnopolsky et al., 1998; Kurokawa et al., 2016). On ancient Mars, a more likely lower limit on *f* is given by the Jeans escape-limited case, where the escape step has $f_{Jeans} \approx 0.26$, resulting in a net fractionation factor *f* = 0.105 (Kurokawa et al., 2016). On early Mars, enhanced UV radiation could make thermal escape to space efficient enough that all H and D atoms escape the exobase, such that the process is non-fractionating. Here, H loss would be limited by diffusion of H from the homopause to the exobase. In this diffusion-limited regime, the diffusion step has $f_{diff} \approx 0.70\text{-}0.81$, resulting in a maximum net fractionation factor $f = 0.33$ (Kurokawa et al., 2016). More fractionating escape (lower $f$) implies less water loss, and thus a smaller share of historical O loss that must be attributed to water, which may instead be attributed to $CO_2$ (Figure 1). Therefore, for estimating upper limit paleo-$pCO_2$, water escape is modeled here with a full range of $f$ values extending to the most fractionating possible value of 0.016. The upper limit on $f$ is 0.33 for diffusion-limited escape. A uniform distribution of $f$ values is assumed between these values (Table 2). We adopt a uniform distribution because we think that the true average value for $f$ is relatively large for the escaped-to-space H reservoir. This is an appropriate assumption because strong solar UV early in Martian history would favor diffusion-limited H escape, at a time when the hydrosphere was largest and thus when most H escape was occurring.

The size of the modern water reservoir on Mars, $R_{t2}$, is given as global equivalent layer of water (GEL) in units of meters (m); the liquid water depth if the reservoir were evenly distributed across the Martian surface. Carr and Head (2015) estimate of $R_{t2} = 34$ m GEL; comprised of 22 m inferred volume shared between polar ice deposits, ≥7 m of ground ice between 50-60° latitude, and <2.6 m of shallow buried ice between 30-60° latitude. The selection by Carr and Head (2015) of 34 m GEL total exceeds the strict sum to ≈ 31.5 m GEL of the above values. An overestimation of $R_{t2}$ propagates to an overestimation of $R_{t1}$ and thus water loss. We use $R_{t2} = 31.5 \pm 1.25$ (1σ) m GEL to encompass the Carr and Head (2015) value, as well as lower estimates of 29 m if shallow buried ice is negligible (Table 2).

The value of $I_{t1}$ comes from measurements of D/H in clay mineral-bound water in 3.5 Ga Yellowknife Bay mudstone (Mahaffy et al., 2015), with a value of $(3.0 \pm 0.2, 2σ)$ x SMOW, where SMOW is the D/H ratio of terrestrial Standard Mean Ocean Water (D/H$_{SMOW}$ = 1.558 x 10$^{-4}$). Modern D/H ratios on Mars vary significantly depending on the measurement technique applied. A measurement of Martian atmospheric D/H at Gale Crater made with the Tunable Laser Spectrometer (TLS) onboard the Curiosity Rover gave $I_{t2}$ = (6 ± 1, 2σ) x SMOW (Webster et al., 2013), consistent with D/H measurements of young indigenous water in Martian meteorites (Greenwood et al., 2008). However, globally averaged measurements of Mars made with spectrometers at Earth-based observatories estimate 7 x SMOW for typical Martian atmosphere, resulting in an inferred D/H ≈ 8 x SMOW for the ice reservoirs where most Martian water resides (Villanueva et al., 2015). Discrepant estimates for $I_{t2}$ propagate to tens of m GEL differences in 3.5 Gyr water loss, or hundreds of mbar $CO_2$-equivalent O loss; thus these differences are important. Direct measurement of D/H in Mars ice by a polar-cap lander could remove this uncertainty (Vos et al., 2019). We run models with both $I_{t2}$ estimates and compare the resulting effects on paleo-$pCO_2$ estimates (Table 3; Figures 3, 4,). Villanueva et al. (2015) do not report an analytical uncertainty for their estimate, so here the range (8.0 ± 1.0, 2σ) x SMOW is applied to compare on equal terms with Webster et al. (2013).

The H loss model in Equation 2 prescribes a single Martian water reservoir that exchanges with the atmosphere rapidly relative to timescales of atmospheric loss. However, alternative H escape models exist to explain intermediate-D/H components in the bound water of SNC meteorites (e.g. Kurokawa et al., 2016). The alternative models invoke an 'unexchangeable' buried ice reservoir which cannot interact directly with the atmosphere, but does supply the exchangeable surface hydrosphere/cryosphere on Mars. Because the surface exchangeable reservoir is constantly supplied with comparatively low-D/H water throughout atmospheric escape, such models generally require greater total water loss to satisfy a given modern D/H constraint. In a similar manner, late (post 3.5 Ga) degassing of juvenile, unfractionated Martian water to the surface would act to buffer the D/H evolution of the hydrosphere (e.g. see Alsaeed and Jakosky, 2019), essentially creating the need for even greater H loss to match D/H data, in only a fraction of the time after 3.5 Ga and thus under weaker solar forcing for escape processes. Moreover, as stated above, we do not consider additional sinks of water that we expect to be unfractionating.

Therefore, without ruling on the plausibility of any of these more complex models for Martian hydrospheric evolution, they are not included in this study because it would imply a significantly greater O loss fraction that must be attributed to water loss, and thus act to drive down paleo-pCO$_2$ estimates (Figure 1), only strengthening our upper limit on post-3.5 Ga pCO$_2$.

*2.6. Integrated model*

Post 3.5 Ga Martian volatile histories were modeled in a Monte Carlo simulation run in MATLAB. Input parameters and their uncertainty distributions are given in Table 2. Dissociative recombination rates through time (Appendix) were calculated $10^5$ times from parameters randomly generated within the uncertainty distributions of the input parameters, and total O loss rates through time for each simulation were calculated as the sum of a given dissociative recombination history and the functions for ion and sputtering loss. Total 3.5 Ga O loss was determined by discrete numerical integration with 1 Myr time steps. The results of O loss scenarios through time for different *γ* (solar EUV dependence of O loss, see Appendix) and *β* (solar EUV devolution parameter, see Appendix) ranges are shown with uncertainty distributions for dissociative recombination in Figure 2.

D/H evolution scenarios were also modelled in a Monte Carlo simulation, with input parameters and their uncertainty distributions given in Table 2. Post-3.5 Ga water loss was calculated by rearranging Equation 6 to give $R_{t1}$, and subtracting $R_{t2}$ from the result. Resulting model distributions for lost water for the local MSL and global telescope-based measurement for $I_{t2}$ are shown in Figure 3. Results were converted to H atom equivalents and combined with O loss results and oxidative sink estimates, in Equation 1, to generate a large number of post-3.5 Ga C (CO$_2$) loss histories. These output distributions were added to integrated CO$_2^+$ ion loss, and the modern atmospheric CO$_2$ reservoirs (6 mbar in atmosphere, 6 mbar in CO$_2$ ice; Putzig et al., 2018) to give probability distributions for 3.5 Ga Martian paleo-pCO$_2$.

The pCO$_2$ values we calculate by applying Equation 1 can take negative or positive values, because total O loss is calculated by integrating the extrapolated model O loss rate backwards in time, and the O loss model is independent of sedimentary O sink and D/H ratio H loss constraints. Negative 3.5 Ga ago pCO$_2$ outputs of our models should not, therefore, be taken to indicate that there was a net source of CO$_2$ to Mars since that time. Rather, negative 3.5 Ga ago pCO$_2$ estimates in our models would indicate that all sinks for O are required to balance H loss to space implied by D/H evolution. In these cases, negligible loss of CO$_2$ would be allowed, such that loss to space would not account for the depletion of a thicker CO$_2$ atmosphere in the past.

## 3. Results

Water loss scenarios are shown in Figure 3, and range up to ≈140 m GEL. The median estimate for water loss was 41 m GEL, or 71 m GEL, depending on the modern D/H constraint, with the estimate of 8 x SMOW instead of 6 x SMOW giving an additional ≈30 m GEL average water loss.

Combinations of three different values for $\gamma$ and two different ranges of $\beta$ in dissociative recombination (Appendix) generated six post-3.5 Ga O loss scenarios (Figure 2), which are consistent with previous estimates (*e.g.* Chassefière et al., 2013; Dong et al., 2018; Lillis et al., 2017). Almost two orders of magnitude variation at the 95% confidence interval in the dissociative recombination loss rate at 3.5 Ga for a given scenario, emphasizes the uncertain nature of extrapolating Martian O loss. When $\gamma$ = 2.64 (Lillis et al., 2017) and $\beta$ from Tu et al. (2015) were used in combination, the 95% upper limit for the 3.5 Ga O loss rate is >10$^{30}$ atoms/s, but in such high EUV scenarios the extrapolated positive relationship between EUV and O loss might break down (R.J. Lillis, personal communication).

Scenarios with all $γ$, $β$, and $I_{t2}$ (modern D/H) input combinations generated 12 model probability distributions (Table 3; Figure 4) for 3.5 Ga paleo-$pCO_2$, which all have negative central estimates. The average additional water loss of ≈30 m GEL in the $I_{t2}$ = 8 x SMOW model runs equates to a further ≈ 1.3 bar equivalent of $CO_2$ that cannot be allocated to the past atmospheric inventory. Dividing the models into two groups based on the modern D/H constraint used, there is still significant variation, which we describe in detail for $I_{t2}$ = 6 x SMOW (it is similar with a constant offset in the 8 x SMOW case).

Central estimates of paleo-$pCO_2$ range from ≈-1290 to -370 mbar (Table 3, Figure 4). Models with $γ$ = 2.64 give strongly positive 95% upper limits for paleo-$pCO_2$, from >4 to several tens of bars (Table 3), which all exceed the independent $pCO_2$ constraints from crater size distributions (Kite et al., 2014; Warren et al., 2019). For other values of $γ$, the only model which results in positive paleo-$pCO_2$ is the one that uses Tu et al. (2015)'s $β$ range and $γ$ = 1.7. All other scenarios give negative paleo-$pCO_2$ at the 95% level between -580 and -120 mbar. These results emphasize a need for a reliable consensus on values for $γ$ and $β$ and their range of uncertainty (Appendix). Alternatively, we can determine what fraction of model runs give a physically acceptable result of paleo-$pCO_2$ >12 mbar, which is larger than the modern $CO_2$ reservoir size and thus is consistent with net loss of $CO_2$ since 3.5 Ga (Figure 4C, D). These values range from ≈0.001 to 42% of model runs (Table 3), with scenarios that favor greater O loss giving a larger proportion of positive paleo-$pCO_2$ results, including values exceeding independent upper limits (e.g. Kite et al., 2014).

## 4. Discussion

Our integrated model includes the best available constraints on Mars atmosphere/hydrosphere evolution. Our central estimates of 3.5 Ga Martian $pCO_2$ are negative. This is physically unrealistic. Therefore, either an important term is missing from our model, or one of our terms must take a value near the edge of its uncertainty range. Consistently negative central estimates of 3.5 Ga Martian $pCO_2$ can be interpreted in two general ways:

i.) Negative paleo-$pCO_2$ could be evidence for a missing sink of Mars' 3.5 Ga surface volatiles. This is because negative $pCO_2$ estimates indicate that MAVEN-derived historical O loss estimates are unable to even fully account for the central estimate for Martian water loss implied by D/H ratio evolution. This observation by itself favors more fractionating (lower $f$) H escape than the diffusion limited endmember case through the last 3.5 Ga of Mars history. However, negative $pCO_2$ estimates also indicate that it is highly unlikely that any significant C loss to space has occurred via an additional unknown channel not seen in the modern day, leaving only the few mbar levels of C escape allowed by extrapolation of modern measurements (Cui et al., 2019; Dong et al., 2018). The required missing sink could be a $CO_2$ reservoir fixed at or below the Martian surface, such as voluminous deposits of $CO_2$ ice, liquid, or clathrates; or $CO_2$ adsorbed on dust or carbonates (Jakosky, 2019; Kurahashi-Nakamura and Tajika, 2006; Manning et al., 2019; Niles et al., 2013). Alternatively, an increase in the surface oxidized sink increases the upper paleo-$pCO_2$ limit for a given O loss to space simulation, per Equation 1. Therefore, the missing volatile sink could also be oxidized lithological unit(s) not counted within the post-3.5 Ga soil and layered sediments.

ii.) If no large missing volatile sinks exist, the accounting of O atoms in the model is complete, and therefore model runs with positive $pCO_2$ outputs are the only ones with physically reasonable combinations of input parameters. Posterior distributions generated from the positive values of the Monte Carlo simulations should provide the true estimate of 3.5 Ga Martian $pCO_2$. A 3.5 Ga $pCO_2$ ≥0.25 bar was possible, but unlikely; particularly considering that the model runs in our study which give the most positive $pCO_2$ results used input parameters have been revised downward with larger datasets. This interpretation also requires a major C/CO escape channel not recorded by MAVEN.

*4.1. Missing volatile sinks*

*4.1.1. CO$_2$-fixing sinks*

Fixing CO$_2$ in (sub)surface reservoirs on Mars is a direct alternative to CO$_2$ loss to space. The negative central estimates of Martian paleo-pCO$_2$ from O and H loss models could be explained by sequestration of a few bars CO$_2$ into post-3.5 Ga carbonate rocks. Carbonate in Martian meteorites (Niles et al., 2013, and references therein), and observations of carbonate on the Martian surface (Boynton et al., 2009; Edwards and Ehlmann, 2015; Ehlmann and Edwards, 2014), show that these phases have formed on Mars, but their role as a globally significant post-3.5 Ga sink is uncertain. Most observed Martian carbonate occurrences suggest local alteration at low water to rock ratios (Niles et al., 2013). Edwards and Ehlmann (2015) found only minor late Noachian-early Hesperian carbonate alteration, equivalent to ≤12 mbar CO$_2$ drawdown, had occurred through basalt alteration in the Nilli Fossae region, suggesting post-Noachian carbonate sequestration was minor. Jakosky and Edwards (2018) estimated <50 mbar of CO$_2$-equivalent Martian carbonate globally. Niles et al. (2013) determined that with a generous crustal carbonate fraction of 0.5 wt% within the top 1 km of Martian crust, ≤0.25 bar of CO$_2$ could be sequestered in the Martian subsurface. Most of this crustal carbonate would have formed long prior to 3.5 Ga, so this CO$_2$ would not have been in the atmosphere when later lakes and rivers were forming. However, global post-3.5 Ga soil and layered sedimentary deposits used in our oxidative weathering balance may also have sequestered CO$_2$. Taking the total mass of these deposits from Supplementary Table 1, and an optimistic 4 wt% of carbonate minerals (Boynton et al., 2009), present as magnesite (MgCO$_3$), we estimate a maximum ≈ 41 mbar CO$_2$-equivalent of carbonate that may be incorporated in the post-3.5 Ga soil and layered sedimentary reservoir. Again, whilst this upper limit value is several times the current exchangeable CO$_2$ inventory on Mars, it falls far short of overcoming the ≈ 1-2 bar deficit to explain both negative pCO$_2$ estimates in this study and additional climatic requirements. Frozen CO$_2$ may have been in sequestered in large volumes deep in the Martian regolith as a result of basal melting of CO$_2$ ice in the polar caps in earlier periods with greater geothermal heat flow (Manning et al., 2019). Indeed, a subsurface radar anomaly exists in an area of past ice-cap basal melting (Whitten et al., 2018). We do not explore this intriguing mechanism further here; however, it could provide a substantial missing C sink in the context of our modeling.

*4.1.2. Oxidation of post-Noachian lava*

The oxidative sink included in the model only accounted for observed layered sedimentary deposits and soil. However, buried paleo-weathering horizons (paleosols) on post-3.5 Ga lava flows on the Tharsis Plateau, Olympus Mons, Elysium, and other young flood lavas, could also sequester O. Surface cratering ages for Tharsis (Tanaka et al., 2014), and the <1.3 Ga crystallization of Shergottite, Nakhlite and Chassignite (SNC) meteorites imply that substantial volumes of lava erupted after the Noachian, and therefore fresh lava may have been periodically exposed, weathered, and then buried by more lava during the 3.5 Ga of most recent Martian history. Kite et al. (2009) estimate a maximum of $5.7 \times 10^{19}$ kg of Martian lava was erupted late, and some fraction of this lava could have been weathered since 3.5 Ga.

Lava flow paleosol development is illustrated in Figure 5. Downward propagation of the weathering front into Martian lava flows can be modelled as a diffusive process, with the depth $L$ of the regolith scaling as $L \approx \sqrt{(\kappa \tau)}$, where $\kappa$ is the diffusivity and $\tau$ is time. Assessment of the NASA InSight landing site found the regolith to be 3-17 m thick in a surface with 1.7 Ga cratering age (Warner et al., 2017), which results in estimates for $\kappa$ in the range $5.3 \times 10^{-9} – 1.7 \times 10^{-7}$ m$^2$/yr. If the weathering front moves downwards as $\sqrt{time}$, lava flow oxidation is maximal for evenly time-spaced eruption of lava flows over 3.5 Gyr (Figure 5) to build up a 10 km characteristic thickness for parts of Tharsis. Regolith on each individual flow deepens to $L_{lava} = \sqrt{(\kappa \tau_{lava})}$ where $\tau_{lava}$ is the time between lava flows. Thinner lava flows enable the most extensive total oxidation, with altimetry estimates placing the lower bound for lava flow thickness on Tharsis at 4 m (Basilevskaya et al., 2006). Assuming oxidation of 39% of total Fe to Fe$^{3+}$ within the developed regolith, as per the global soil layer (Morris et al., 2006), the maximum O sink available from young lava paleosol oxidation is ≈3.4 x $10^{18}$ moles of O, which equates to 0.42 m GEL water or 19.1 mbar $CO_2$ equivalent oxygen. The minimum oxidation scenario, instantaneous eruption of all lava and weathering at a single surface for 3.5 Gyr, sequesters ≤$10^{17}$ moles of O, equating to 0.008 m GEL water or 0.38 mbar $CO_2$. Greater Fe oxidation per unit mass, or faster diffusive growth of regolith on a younger Mars, could increase this number, but not to the ≈1-2 bar $CO_2$ equivalent required.

### *4.2. Posterior distributions for positive $pCO_2$*

Posterior distributions for paleo-$pCO_2$ were generated by considering only the results with $pCO_2$ >12 mbar at 3.5 Ga (Figure 4C, D). These represent scenarios with net $CO_2$ loss to space since 3.5 Ga. The probability distributions cut off below 12 mbar are shown in panels C and D of Figure 4. If large missing volatile sinks are discounted, results above 250 mbar, the lowest $pCO_2$ resulting in runoff that has been used in Early Mars climate models (Urata and Toon, 2013), would be expected to be consistent with lake forming climates.

The proportion of runs with paleo-$pCO_2$>12 mbar is 0.001 to 42 % (Table 3). Scenarios with identical O loss contain a smaller proportion of results with positive $pCO_2$ when the 8 x SMOW modern D/H constraint (Villanueva et al., 2015) is used. However, in every case, the more enriched modern D/H also gives higher upper limits on pCO2 at the 95% level in the posterior distribution. These distributions were cut off further from the central value in the original simulations, giving increased importance to their long tails (Figure 4D).

Model runs using $γ$ = 2.64 (Lillis et al., 2017), result in very high upper limit positive paleo-$pCO_2$ estimates at the 95% level. As we have outlined above, this high value for $γ$ is probably an overestimate of the true value. However, results of models run with modest O loss rates resulting from combinations of lower values of $γ$ and $β$, still result in upper limits on paleo-$pCO_2$ of ≥1.3 bar (Table 3). Three error-propagation scenarios; two using 6 x SMOW and one using 8 x SMOW for the modern D/H constraint; which also made use of more conservative ranges for both $β$ and $γ$, had positive pCO$_2$ distributions for which the posterior 95% percentiles (1.3-2.7 bar) were consistent with independent upper limits imposed by crater size distributions (Kite et al., 2014). Posterior distributions generated from combined O and H loss simulations cannot in any case, rule out the >1 bar pCO$_2$ atmospheres required by many lake-forming climate solutions at the 95% confidence level. This suggests that existing climate models for Mars can be consistent with a subset of volatile evolution histories on Mars, with the caveat that these positive outcomes are unlikely results of our model overall. Results with pCO$_2$ >0.25 bar in all posterior distributions for 3.5 Ga pCO$_2$ are consistent with the minimum constraint required by the climate model of Urata and Toon (2013), suggesting that low probability outcomes of even the most pessimistic estimates of the Mars volatile budget could enable lake-forming climates under at least some climate models (Kite et al., 2017; Mansfield et al., 2018; Ramirez et al., 2014). These hypotheses are potentially testable via *Curiosity* measurements of $δ^{13}$C in the rocks of Mt. Sharp (Franz et al. 2015).

## 5. Conclusions

Mars no longer has the thick CO$_2$ atmosphere that climate models indicate would have been required to sustain lakes that formed more recently than 3.5 Ga ago. This suggests that large quantities of C have been lost from Mars's atmosphere. Combining multiple independent constraints on post-3.5 Ga evolution of Martian volatile reservoirs (Figure 1) has enabled us address whether this C was lost to space.. Our key results are illustrated in Figure 6. In order to address this problem, we estimated both the ranges of Martian oxygen and hydrogen loss to space over the past 3.5 Gyr using an error propagation approach (Figures 2-4), breaking a previous degeneracy in attributing oxygen lost to space to either H$_2$O or CO$_2$. We also refined the picture of the Martian oxidative soil and layered sediment oxygen sink (Figure 5), finding it to be small relative to the flux of oxygen lost to space. Despite the reddening of Mars attributable to surficial oxidation of iron equating to <0.5m GEL of H$_2$O-equivalent oxygen, whereas the main oxygen-consuming component in the oxidative soil and layered sedimentary budget could have been sulfur, provided that volcanic sulfur gases on Mars reflect the redox state of magma inferred from meteorites. With our multiproxy approach we determined upper limits on 3.5 Ga Martian paleo-$pCO_2$ that were agnostic about loss mechanisms for carbon itself. Despite significant spread in results, a consistent feature of our models was negative (-370 to -2660 mbar) central estimates for Martian 3.5 Ga paleo-$pCO_2$, with the least negative results only being obtained with model parameters which do not represent the best current understanding of atmospheric oxygen escape.

Our negative central $pCO_2$ estimates suggest either i.) a missing component of the volatile loss model, such as a large, unaccounted for volatile sink on Mars; ii.) that post-3.5 Ga atmospheric loss processes were more vigorous than currently thought. Evidence suggests post-3.5 Ga carbonate formation was minor, and a previously unconsidered oxygen sink of $Fe^{3+}$-rich paleosols on Tharsis lavas can only account for a small fraction of the $CO_2$-equivalent oxygen needed for >250 mbar atmospheres postulated to be required by climate models. More likely, our results indicate that of C loss to space cannot account for loss of hundreds of mbar of $CO_2$ in the last 3.5 Ga of Martian history. The apparent lack of sufficient sinks for C lost since 3.5 Ga ago could be solved by the discovery of large, as-yet undiscovered, carbon sinks in Martian (sub)surface deposits. Our preferred solution, missing C sinks, is only one of three possible solutions to this problem (Figure 6). These three possible solution are that i.) there is a missing C sink on (in) Mars's (sub)surface; ii.) climate solutions for sustained runoff on Mars with atmospheric $CO_2$ pressure <<250 mbar; or iii.) crater chronologies on Mars being sufficiently incorrect that all apparently young rivers and lakes date are actually older than 3.5 Ga (Figure 6). Each of these possibilities merits substantial dedicated investigation in future studies.

**Acknowledgements**

We thank Rob Lillis for insightful discussions that significantly improved the manuscript. We also thank Bruce Jakosky and an anonymous reviewer for thorough and thoughtful reviews, and Itay Halevy for editorial handling of the manuscript. A. W. Heard was supported by NASA grant NNH16ZDA001N to Nicolas Dauphas and A. W. Heard. E. S. Kite was supported by NASA grant NNX16AG55G to E. S. Kite.

**Figure Captions**

**Figure 1 –** A.) Sketch of Martian H, C, and O fluxes considered in our study. C species can also be lost to space, either as $CO_2$ molecules or as constituent C, O, and CO. Atmospheric $CO_2$ might be sequestered into (sub)surface carbonates, $CO_2$ liquids and ices, and clathrates. Atmospheric $H_2O$ is sourced from polar caps and ground ice, and H and O can be lost to space. Both $CO_2$- and $H_2O$-derived O not lost to space can be sequestered in soils and layered sedimentary rocks during oxidative weathering processes, principally through interactions with reduced Fe and S species. B.) Graphical summary of the modelling approach and relative sizes of various O sinks as identified in our study. The heavy black vertical line separates the state of knowledge before this study (to the left), from the stronger constraint placed on $pCO_2$ by considering the D/H constraint on $H_2O$ loss, to the right.

**Figure 2 –** Martian oxygen loss over time for different EUV flux dependence ($\gamma_{2.64}$, $\gamma_{1.7}$, $\gamma_{1.0}$) and solar EUV evolution ($\beta_{rot}$, $\beta_{lum}$) scenarios. A.) $\gamma_{2.64}$, $\beta_{rot}$; B.) $\gamma_{1.7}$, $\beta_{rot}$; C.) $\gamma_{1.0}$, $\beta_{rot}$; D.) $\gamma_{2.64}$, $\beta_{lum}$; E.) $\gamma_{1.7}$, $\beta_{lum}$; F.) $\gamma_{1.0}$, $\beta_{lum}$. Refer to Table 2 and Appendix Equaitions A1 and A2, for these parameter values and sources. The heavy black line shows the evolution of the dissociative recombination (DR) loss flux: dark and light grey shaded

areas constrain the 1σ and 2σ uncertainty envelope on this flux, respectively. The red solid, dot-dashed, and dashed lines show the loss fluxes of pickup $O^+$ ions, sputtered O atoms, and $O^+$ ionospheric outflow, respectively. Functional forms follow Chassefière et al., (2013), adjusted to fit updated modern loss fluxes recorded by MAVEN (Jakosky et al., 2018). The blue solid line shows the $CO_2^+$ ion loss channel, modeled following Dong et al., (2018). The orange bar indicates the full range of observed hot O (dissociative recombination) loss rates inferred over Mars observational history, spanning $5 \times 10^{24} – 4 \times 10^{26}$ atoms/s. The right-hand axis shows the loss rates in $pCO_2$ equivalents in mbar/Gyr.

**Figure 3** – Probability distributions of water loss simulations since 3.5 Ga for Mars, in terms of meters Global Equivalent Layer of water, constrained by D/H ratios. The two scenarios use identical parameter ranges for D/H at 3.5 Ga, the size of the modern Martian water reservoir, and the fractionation factor, but with different estimates for the average modern D/H of water on Mars (Villanueva et al., 2015; Webster et al., 2013). The difference between central estimates is ≈ 30 m GEL, equating to a little over 1 bar $CO_2$ equivalent O.

**Figure 4** – Probability distributions of 3.5 Ga paleo-$pCO_2$, for different O loss scenarios and utilizing water loss scenarios constrained by A.) the 6 x SMOW and, B.) 8 x SMOW modern D/H ratios (Villanueva et al., 2015; Webster et al., 2013). The light black vertical line in each case shows the minimum $pCO_2$ of 250 mbar shown to enable 3.5 Ga runoff in climate models (Urata and Toon, 2013). C.) and D.) show the part of the probability distributions of 3.5 Ga paleo-$pCO_2$ with results >12 mbar (i.e., indicating net loss of $CO_2$ since 3.5 Ga), from A.) and B.) respectively. The percentage of simulations falling above 12 mbar for each simulation can be found in Table 3. The black arrow indicates model paleo-$pCO_2$ results greater than 5 bar. These constitute <5 % of all simulations except for the case of $\gamma_{2.64}$ combined with $\beta_{rot}$, where they constitute <20 % of all simulations.

**Figure 5** – Cartoon illustrating scenarios for the formation of oxidized paleosols on lava flows within a ≈ 10 km thick column of lava flows in the Tharsis region, taken as representative for young post-Noachian lavas on Mars. Regolith develops on a fresh lava surface as a diffusive process, deepening as the square root of the exposure time. The optimum scenario for oxidation is for uniform periodic eruption of the thinnest lava flows possible (>4 m; Basilevskaya et al., 2006), and development of a regolith on top of each one. The worst-case scenario for oxidation is for geologically instantaneous eruption of the entire post-3.5 Ga Tharsis column, and oxidation at a single surface for 3.5 Gyr. Assuming oxidation of Fe to a degree consistent with typical Martian soil (Morris et al., 2006), neither scenario can sequester enough O atoms to explain a large fraction of Mars' post-3.5 Ga $H_2O$ loss of presumed post-3.5 Ga $CO_2$ loss.

**Figure 6** – Graphical illustration of new constraints and future directions for investigation provided by this study. A. Geologic data suggest lakes persisted after almost all atmosphere was lost: Evidence for lakes and rivers up to 1-2 Ga (Dickson et al., 2009) and abundant evidence in the rock record (including lacustrine mudstones) at 3.5 Ga require higher $pCO_2$ (according to existing climate models), than that provided by known $CO_2$ sinks, or by unknown C loss mechanisms implied by O sinks. B. Possible ways to reconcile negligible $CO_2$ loss to space with rivers and lakes after 3.5 Ga ago are large missing C sinks; new climate models capable of producing extensive runoff at far lower $pCO_2$ than existing models; and revision of age data such that all evidence for lakes and rivers could predate loss of major C sinks;.

**Appendix - Oxygen loss to space**

*Major O loss process: Oxygen loss through dissociative recombination*

Careful extrapolation of photochemical O loss through time is central to determining a reasonable spread of Martian volatile budgets back to 3.5 Ga. Table 2 contains a list of the parameters used in our modeling. Photochemical loss is driven by solar extreme ultraviolet (EUV, 10-92 nm) radiation. MAVEN constrained a temporal spread in dissociative recombination O loss driven primarily by EUV variation associated with the solar cycle. A number of approaches have yielded average determinations on the order of 5 x $10^{25}$ O $s^{-1}$, and here we apply the recommended range of Lillis et al. (2017), with a mean of 4.3 x $10^{25}$ O $s^{-1}$, and with upper and lower (1σ) bounds of 9.6 x $10^{25}$ O $s^{-1}$ and 1.9 x $10^{25}$ O $s^{-1}$ respectively, following a log-normal distribution (Table 2). The central value reported by Lillis et al. is in flux (R. J. Lillis, personal communication), therefore including this range should adequately encompass temporal variation in the loss rate, including short-term solar storm enhancement of the EUV flux which can enhance photochemical loss by a factor of approximately 10.

Dissociative recombination O loss flux *F*, is assumed to follow a power law relationship with the EUV ionization frequency *I*

$$F = BI^\gamma \qquad (A1)$$

with the power law exponent *γ*, where *B* is a fitting parameter. Published literature indicates *γ* = 2.64 ± 0.60 (1σ) (Lillis et al., 2017). However, an enlarged dataset suggests the average value of *γ* should be revised downward to *γ* = 1.7 ± 0.39 (R. J. Lillis, personal communication, where we have assumed identical proportional size of error bars to Lillis *et al.,* 2017). A simple analytical theory suggests dissociative recombination is directly proportional to the EUV irradiance, with *γ* = 1 (Cravens et al., 2017). We run models with all three values and compare the results (Table 2, Figure 2).

Extrapolation of the loss flux backwards in time requires an expression for the time evolution of the ionization frequency

$$I_{\text{past}} = I_{\text{present}} \left(\frac{t_{\text{past}}}{4.5 \text{ Gyr}}\right)^{-\beta} \quad \text{(A2)}$$

where $t_{\text{past}}$ is the age of Mars at some point in the past in Gyr, and $\beta$ is the power law exponent for the decay of solar EUV intensity over time, with the form $t^{-\beta}$, where t is the stellar age (Lillis et al., 2017). Converting flux $F$ from (2) to planetary loss rate $R$ through multiplying by the surface area of Mars, recasting in terms of Ga and combining Equation 1 and Equation 2 yields the expression for the past dissociative recombination flux $R_{\text{past}}$ at a given time before present $t_{\text{Ga}}$

$$R_{\text{past}} = R_{\text{present}} \left(\frac{4.5}{4.5 - t_{\text{Ga}}}\right)^{\gamma\beta} \quad \text{(A3)}$$

where here the maximum value of $t_{\text{Ga}}$ is 3.5 (Lillis et al., 2017).

The value of $\beta$ (Table 2) is uncertain in very young stars because stellar X-Ray and EUV radiation scale with magnetic activity inside stars, which depends on the stellar dynamo and thus the star's rotation (Tu et al., 2015). Young stars show a range of rotational speeds but by the 1 Gyr age of the Sun when our models begin, such rotational activity converges on a narrower range because fast-spinning stars spin down faster (Mamajek and Hillenbrand, 2008). The solar EUV evolution exponent $\beta$ can be determined by measurement of UV and X-ray luminosities for stars of known age (with a fitted relationship between X-ray and EUV luminosity), or through modeling the rotational evolution of stars and calculating the anticipated X-ray (and consequently, fitted EUV), luminosities (Ribas et al., 2005; Sanz-Forcada et al., 2011; Tu et al., 2015). Ribas et al. (2005) found that the EUV decay power law exponent $\beta$ = 1.20 or 1.23 within a sample of 6 G stars when spectra were averaged over the wavelength ranges 10-36 and 0.1-110 nm respectively. Sanz-Forcada et al. (2011) examined 29 M to F class stars (3 M, 26 FGK) with exoplanetary systems and found $\beta$ = 1.24 ± 0.15 for EUV luminosity. Tu et al. (2015) used hundreds of observations of young stars combined with a stellar evolution model and determined a spread in early X-Ray and EUV luminosities with $β$ = 1.22 with large and highly asymmetric uncertainty, with 10th and 90th percentiles of 0.96 and 2.15, respectively. While the study of Ribas et al. (2005) has a smaller sample size, the parameter $β$ in that study is constrained by stars spanning 6.7 Gyr in age with 3 of 6 samples being older than 1 Gyr. The study of Sanz-Forcada et al. (2011) includes a fraction of stars up to 3 Gyr in age, and Tu et al. (2015) exclusively made use of stars less than 600 million years (Myr) old, and extrapolated their model through to the age of the Sun. Those latter two studies do not utilize direct observations in the EUV spectral range, but instead apply a fitted relationship between X-Ray and EUV luminosity given by Sanz-Forcada et al. (2011)

$$\log L_{\text{EUV}} = 4.8 + 0.86 \log L_{\text{X}} \quad \text{(A4)}$$

where $L_{\text{EUV}}$ and $L_{\text{X}}$ are the modeled EUV and observed X-Ray luminosities respectively.

Early rotational evolution of stars leads to a spread in their X-Ray (and inferred EUV) luminosities before 500 Myr, after which these parameters converge on a more limited spread of values for a given stellar age (Tu et al., 2015). The implication of this is that Tu et al. (2015)'s confidence interval, defined as it is by young and highly rotationally variable stars, overestimates the level of uncertainty in the value of $β$ for later stages of stellar evolution. In our work we are interested in the evolution of a >1 Gyr old star, the post-3.5 Ga Sun, and its influence on volatile evolution on Mars For this age range, $β$ shows less spread. A larger upper limit value of $β$ corresponds to historically higher EUV flux and thus enhanced early photochemical loss of O (Lillis et al., 2017). Determination of conservative upper limits for the total O loss will result in robust upper limits on paleo-$pCO_2$ per Equation 1. Sensitivity of Mars's photochemical O loss history to the EUV evolution of the Sun is assessed by using two ranges of $β$ (Table 2), from Sanz-Forcada *et al.* (2011) and Tu et al. (2015), and comparing results.

*Smaller O loss processes: Oxygen ion and sputtering loss*

Other O loss channels are also significant for Martian volatile history (Figure 2). Following the approach of Jakosky et al. (2018), we include two additional O loss channels; O ion loss, and sputtering O loss. Previous workers such as Chassefière et al. (2013) considered the separate O ion loss channels of pickup ion loss and ionospheric outflow. However, Jakosky et al. (2018) reported a single number for O ion loss of $5 \times 10^{24}$ $O^+$ $s^{-1}$. In detail, ionospheric outflow loss is smaller (by $\approx 10^2$ times) than pickup ion loss, and barely affects integrated O loss (Chassefière et al., 2013). However, the unmodified function from Chassefière et al. (2013) is included here for completeness. Pickup ion loss at present day is fixed at the modern ion loss rates determined by MAVEN. Following Jakosky et al. (2018), the evolution of the loss flux back in time is modeled following the functional form of Chassefière et al. (2013), after Lammer et al. (2003).

The functions chosen in this study for the ion and sputtering loss channels give higher loss rates at 3.5 Ga than another recent, MAVEN-led study of Dong et al. (2018). These parameters are chosen to allow the most generous upper limits on historical Martian O loss and thus paleo-$pCO_2$. In the work of Dong et al. (2018), total O ion loss at 3.5 Ga is $2.4 \times 10^{26}$ $O^+$ $s^{-1}$, compared to $\approx 1.4 \times 10^{27}$ $O^+$ $s^{-1}$ in the parameterization of Chassefière et al. (2013). Our selection is consistent with the aim of obtaining strong upper limits on paleo-pCO2, without intending to comment on the relative accuracy of these values.

Sputtering loss is driven by the reimpacting of $O^+$ ions, accelerated by the electrical field of the solar wind, back into the atmosphere, where transfer of momentum from these ions to neutrals can remove even heavy atoms. MAVEN does not directly measure sputtering. Sputtering loss is determined through model calculations using measured properties of reimpacting ion fluxes. The inferred average sputtering loss rate from modern Mars is $3 \times 10^{24}$ O $s^{-1}$ (Jakosky et al., 2018). In this study, following Jakosky et al. (2018), we use the functional form from Chassefière et al. (2013) for evolution of the sputtering rate through time, adjusted to fit the modern value.

**References**

Alsaeed, N.R., Jakosky, B.M., 2019. Mars Water and D/H Evolution from 3 Ga to Present. LPI Contrib. 2089, 6059.

Barabash, S., Fedorov, A., Lundin, R., Sauvaud, J.-A., 2007. Martian Atmospheric Erosion Rates. Science 315, 501–503. https://doi.org/10.1126/science.1134358

Basilevskaya, E.A., Neukum, G., the HRSC Co-Investigator Team, 2006. The Olympus volcano on Mars: Geometry and characteristics of lava flows. Sol. Syst. Res. 40, 375–383. https://doi.org/10.1134/S0038094606050029

Boynton, W.V., Ming, D.W., Kounaves, S.P., Young, S.M.M., Arvidson, R.E., Hecht, M.H., Hoffman, J., Niles, P.B., Hamara, D.K., Quinn, R.C., Smith, P.H., Sutter, B., Catling, D.C., Morris, R.V., 2009. Evidence for Calcium Carbonate at the Mars Phoenix Landing Site. Science 325, 61–64. https://doi.org/10.1126/science.1172768

Carr, M.H., Head, J.W., 2015. Martian surface/near-surface water inventory: Sources, sinks, and changes with time. Geophys. Res. Lett. 42, 726–732. https://doi.org/10.1002/2014GL062464

Catling, D.C., Kasting, J.F., 2017. Atmospheric Evolution on Inhabited and Lifeless Worlds.

Chassefière, E., Langlais, B., Quesnel, Y., Leblanc, F., 2013. The fate of early Mars' lost water: The role of serpentinization. J. Geophys. Res. Planets 118, 1123–1134. https://doi.org/10.1002/jgre.20089

Cravens, T.E., Rahmati, A., Fox, J.L., Lillis, R., Bougher, S., Luhmann, J., Sakai, S., Deighan, J., Lee, Y., Combi, M., Jakosky, B., 2017. Hot oxygen escape from Mars: Simple scaling with solar EUV irradiance. J. Geophys. Res. Space Phys. 122, 1102–1116. https://doi.org/10.1002/2016JA023461

Cui, J., Wu, X.-S., Gu, H., Jiang, F.-Y., Wei, Y., 2019. Photochemical escape of atomic C and N on Mars: clues from a multi-instrument MAVEN dataset. Astron. Astrophys. 621, A23. https://doi.org/10.1051/0004-6361/201833749

Dickson, J.L., Fassett, C.I., Head, J.W., 2009. Amazonian-aged fluvial valley systems in a climatic microenvironment on Mars: Melting of ice deposits on the interior of Lyot Crater. Geophys. Res. Lett. 36. https://doi.org/10.1029/2009GL037472

Dong, C., Lee, Y., Ma, Y., Lingam, M., Bougher, S., Luhmann, J., Curry, S., Toth, G., Nagy, A., Tenishev, V., Fang, X., Mitchell, D., Brain, D., Jakosky, B., 2018. Modeling Martian Atmospheric Losses over Time: Implications for Exoplanetary Climate Evolution and Habitability. Astrophys. J. 859, L14. https://doi.org/10.3847/2041-8213/aac489

Edwards, C.S., Ehlmann, B.L., 2015. Carbon sequestration on Mars. Geology 43, 863–866. https://doi.org/10.1130/G36983.1

Ehlmann, B.L., Edwards, C.S., 2014. Mineralogy of the Martian Surface. Annu. Rev. Earth Planet. Sci. 42, 291–315. https://doi.org/10.1146/annurev-earth-060313-055024

Ehlmann, B.L., Mustard, J.F., Murchie, S.L., Bibring, J.-P., Meunier, A., Fraeman, A.A., Langevin, Y., 2011. Subsurface water and clay mineral formation during the early history of Mars. Nature 479, 53–60. https://doi.org/10.1038/nature10582

Franz, H. B., Mahaffy, P. R., Stern, J., Archer, P., Jr., Conrad, P., Eigenbrode, J., Freissinet, C., Glavin, D., Grotzinger, J. P., Jones, J., et al., 2015, ISOTOPIC COMPOSITION OF CARBON DIOXIDE RELEASED FROM CONFIDENCE HILLS SEDIMENT AS MEASURED BY THE SAMPLE ANALYSIS AT MARS (SAM)

QUADRUPOLE MASS SPECTROMETER, Lunar and Planetary Science Conference 2015.Grant, J.A., Wilson, S.A., 2012. A possible synoptic source of water for alluvial fan formation in southern Margaritifer Terra, Mars. Planet. Space Sci., Mars Habitability 72, 44–52. https://doi.org/10.1016/j.pss.2012.05.020

Greenwood, J.P., Itoh, S., Sakamoto, N., Vicenzi, E.P., Yurimoto, H., 2008. Hydrogen isotope evidence for loss of water from Mars through time. Geophys. Res. Lett. 35. https://doi.org/10.1029/2007GL032721

Haberle, R.M., Catling, D.C., Carr, M.H., Zahnle, K.J., 2017. The Early Mars Climate System, in: The Atmosphere and Climate of Mars, Cambridge Planetary Science. Cambridge University Press, Cambridge, pp. 526–568.

Halevy, I., Zuber, M.T., Schrag, D.P., 2007. A Sulfur Dioxide Climate Feedback on Early Mars. Science 318, 1903–1907. https://doi.org/10.1126/science.1147039

Hurowitz, J.A., Fischer, W.W., Tosca, N.J., Milliken, R.E., 2010. Origin of acidic surface waters and the evolution of atmospheric chemistry on early Mars. Nat. Geosci. 3, 323–326. https://doi.org/10.1038/ngeo831

Irwin III, R.P., Lewis, K.W., Howard, A.D., Grant, J.A., 2015. Paleohydrology of Eberswalde crater, Mars. Geomorphology, Planetary Geomorphology: Proceedings of the 45th Annual Binghamton Geomorphology Symposium, held 12-14 September 2014 in Knoxville, Tennessee, USA 240, 83–101. https://doi.org/10.1016/j.geomorph.2014.10.012

Jakosky, B.M., 2019. The $CO_2$ inventory on Mars. Planet. Space Sci. 175, 52–59. https://doi.org/10.1016/j.pss.2019.06.002

Jakosky, B.M., 1991. Mars volatile evolution: Evidence from stable isotopes. Icarus 94, 14–31. https://doi.org/10.1016/0019-1035(91)90138-J

Jakosky, B.M., Brain, D., Chaffin, M., Curry, S., Deighan, J., Grebowsky, J., Halekas, J., Leblanc, F., Lillis, R., Luhmann, J.G., Andersson, L., Andre, N., Andrews, D., Baird, D., Baker, D., Bell, J., Benna, M., Bhattacharyya, D., Bougher, S., Bowers, C., Chamberlin, P., Chaufray, J.-Y., Clarke, J., Collinson, G., Combi, M., Connerney, J., Connour, K., Correira, J., Crabb, K., Crary, F., Cravens, T., Crismani, M., Delory, G., Dewey, R., DiBraccio, G., Dong, C., Dong, Y., Dunn, P., Egan, H., Elrod, M., England, S., Eparvier, F., Ergun, R., Eriksson, A., Esman, T., Espley, J., Evans, S., Fallows, K., Fang, X., Fillingim, M., Flynn, C., Fogle, A., Fowler, C., Fox, J., Fujimoto, M., Garnier, P., Girazian, Z., Groeller, H., Gruesbeck, J., Hamil, O., Hanley, K.G., Hara, T., Harada, Y., Hermann, J., Holmberg, M., Holsclaw, G., Houston, S., Inui, S., Jain, S., Jolitz, R., Kotova, A., Kuroda, T., Larson, D., Lee, Y., Lee, C., Lefevre, F., Lentz, C., Lo, D., Lugo, R., Ma, Y.-J., Mahaffy, P., Marquette, M.L., Matsumoto, Y., Mayyasi, M., Mazelle, C., McClintock, W., McFadden, J., Medvedev, A., Mendillo, M., Meziane, K., Milby, Z., Mitchell, D., Modolo, R., Montmessin, F., Nagy, A., Nakagawa, H., Narvaez, C., Olsen, K., Pawlowski, D., Peterson, W., Rahmati, A., Roeten, K., Romanelli, N., Ruhunusiri, S., Russell, C., Sakai, S., Schneider, N., Seki, K., Sharrar, R., Shaver, S., Siskind, D.E., Slipski, M., Soobiah, Y., Steckiewicz, M., Stevens, M.H., Stewart, I., Stiepen, A., Stone, S., Tenishev, V., Terada, N., Terada, K., Thiemann, E., Tolson, R., Toth, G., Trovato, J., Vogt, M., Weber, T., Withers, P., Xu, S., Yelle, R., Yiğit, E., Zurek, R., 2018. Loss of the Martian atmosphere to space: Present-day loss rates determined from MAVEN observations and integrated loss through time. Icarus. https://doi.org/10.1016/j.icarus.2018.05.030

Jakosky, B.M., Edwards, C.S., 2018. Inventory of CO 2 available for terraforming Mars. Nat. Astron. 2, 634. https://doi.org/10.1038/s41550-018-0529-6

Jakosky, B.M., Phillips, R.J., 2001. Mars' volatile and climate history. Nature 412, 237. https://doi.org/10.1038/35084184

Kite, E.S., 2019. Geologic Constraints on Early Mars Climate. Space Sci. Rev. 215, 10. https://doi.org/10.1007/s11214-018-0575-5

Kite, E.S., Gao, P., Goldblatt, C., Mischna, M.A., Mayer, D.P., Yung, Y.L., 2017. Methane bursts as a trigger for intermittent lake-forming climates on post-Noachian Mars. Nat. Geosci. 10, 737–740. https://doi.org/10.1038/ngeo3033

Kite, E.S., Matsuyama, I., Manga, M., Perron, J.T., Mitrovica, J.X., 2009. True Polar Wander driven by late-stage volcanism and the distribution of paleopolar deposits on Mars. Earth Planet. Sci. Lett. 280, 254–267. https://doi.org/10.1016/j.epsl.2009.01.040

Kite, E.S., Mayer, D.P., Wilson, S.A., Davis, J.M., Lucas, A.S., Quay, G.S. de, 2019. Persistence of intense, climate-driven runoff late in Mars history. Sci. Adv. 5, eaav7710. https://doi.org/10.1126/sciadv.aav7710

Kite, E.S., Williams, J.-P., Lucas, A., Aharonson, O., 2014. Low palaeopressure of the martian atmosphere estimated from the size distribution of ancient craters. Nat. Geosci. 7, 335–339. https://doi.org/10.1038/ngeo2137

Krasnopolsky, V.A., Mumma, M.J., Gladstone, G.R., 1998. Detection of Atomic Deuterium in the Upper Atmosphere of Mars. Science 280, 1576–1580. https://doi.org/10.1126/science.280.5369.1576

Kurahashi-Nakamura, T., Tajika, E., 2006. Atmospheric collapse and transport of carbon dioxide into the subsurface on early Mars. Geophys. Res. Lett. 33. https://doi.org/10.1029/2006GL027170

Kurokawa, H., Sato, M., Ushioda, M., Matsuyama, T., Moriwaki, R., Dohm, J.M., Usui, T., 2014. Evolution of water reservoirs on Mars: Constraints from hydrogen isotopes in martian meteorites. Earth Planet. Sci. Lett. 394, 179–185. https://doi.org/10.1016/j.epsl.2014.03.027

Kurokawa, H., Usui, T., Sato, M., 2016. Interactive evolution of multiple water-ice reservoirs on Mars: Insights from hydrogen isotope compositions. Geochem. J. 50, 67–79. https://doi.org/10.2343/geochemj.2.0407

Lammer, H., Lichtenegger, H.I.M., Kolb, C., Ribas, I., Guinan, E.F., Abart, R., Bauer, S.J., 2003. Loss of water from Mars:: Implications for the oxidation of the soil. Icarus 165, 9–25. https://doi.org/10.1016/S0019-1035(03)00170-2

Lillis, R.J., Deighan, J., Fox, J.L., Bougher, S.W., Lee, Y., Combi, M.R., Cravens, T.E., Rahmati, A., Mahaffy, P.R., Benna, M., Elrod, M.K., McFadden, J.P., Ergun, R.E., Andersson, L., Fowler, C.M., Jakosky, B.M., Thiemann, E., Eparvier, F., Halekas, J.S., Leblanc, F., Chaufray, J.-Y., 2017. Photochemical escape of oxygen from Mars: First results from MAVEN in situ data. J. Geophys. Res. Space Phys. 122, 3815–3836. https://doi.org/10.1002/2016JA023525

Liu, K.-K., Epstein, S., 1984. The hydrogen isotope fractionation between kaolinite and water. Chem. Geol. 46, 335–350. https://doi.org/10.1016/0009-2541(84)90176-1

Mahaffy, P.R., Webster, C.R., Stern, J.C., Brunner, A.E., Atreya, S.K., Conrad, P.G., Domagal-Goldman, S., Eigenbrode, J.L., Flesch, G.J., Christensen, L.E., Franz, H.B., Freissinet, C., Glavin, D.P., Grotzinger, J.P., Jones, J.H., Leshin, L.A., Malespin, C., McAdam, A.C., Ming, D.W., Navarro-Gonzalez, R.,

Niles, P.B., Owen, T., Pavlov, A.A., Steele, A., Trainer, M.G., Williford, K.H., Wray, J.J., Team, the M.S., 2015. The imprint of atmospheric evolution in the D/H of Hesperian clay minerals on Mars. Science 347, 412–414. https://doi.org/10.1126/science.1260291

Mamajek, E.E., Hillenbrand, L.A., 2008. Improved Age Estimation for Solar-Type Dwarfs Using Activity-Rotation Diagnostics. Astrophys. J. 687, 1264–1293. https://doi.org/10.1086/591785

Manning, C.V., Bierson, C., Putzig, N.E., McKay, C.P., 2019. The formation and stability of buried polar CO2 deposits on Mars. Icarus 317, 509–517. https://doi.org/10.1016/j.icarus.2018.07.021

Mansfield, M., Kite, E.S., Mischna, M.A., 2018. Effect of Mars Atmospheric Loss on Snow Melt Potential in a 3.5 Gyr Mars Climate Evolution Model. J. Geophys. Res. Planets 123, 794–806. https://doi.org/10.1002/2017JE005422

Michael, G.G., 2013. Planetary surface dating from crater size–frequency distribution measurements: Multiple resurfacing episodes and differential isochron fitting. Icarus 226, 885–890. https://doi.org/10.1016/j.icarus.2013.07.004

Michalski, J., Niles, P.B., 2012. Atmospheric origin of Martian interior layered deposits: Links to climate change and the global sulfur cycle. Geology 40, 419–422. https://doi.org/10.1130/G32971.1

Morris, R.V., Klingelhöfer, G., Schröder, C., Rodionov, D.S., Yen, A., Ming, D.W., de Souza, P.A., Wdowiak, T., Fleischer, I., Gellert, R., Bernhardt, B., Bonnes, U., Cohen, B.A., Evlanov, E.N., Foh, J., Gütlich, P., Kankeleit, E., McCoy, T., Mittlefehldt, D.W., Renz, F., Schmidt, M.E., Zubkov, B., Squyres, S.W., Arvidson, R.E., 2006. Mössbauer mineralogy of rock, soil, and dust at Meridiani Planum, Mars: Opportunity's journey across sulfate-rich outcrop, basaltic sand and dust, and hematite lag deposits. J. Geophys. Res. Planets 111, E12S15. https://doi.org/10.1029/2006JE002791

Niles, P.B., Catling, D.C., Berger, G., Chassefière, E., Ehlmann, B.L., Michalski, J.R., Morris, R., Ruff, S.W., Sutter, B., 2013. Geochemistry of Carbonates on Mars: Implications for Climate History and Nature of Aqueous Environments. Space Sci. Rev. 174, 301–328. https://doi.org/10.1007/s11214-012-9940-y

Palucis, M.C., Dietrich, W.E., Williams, R.M.E., Hayes, A.G., Parker, T., Sumner, D.Y., Mangold, N., Lewis, K., Newsom, H., 2016. Sequence and relative timing of large lakes in Gale crater (Mars) after the formation of Mount Sharp. J. Geophys. Res. Planets 121, 472–496. https://doi.org/10.1002/2015JE004905

Putzig, N.E., Smith, I.B., Perry, M.R., Foss, F.J., Campbell, B.A., Phillips, R.J., Seu, R., 2018. Three-dimensional radar imaging of structures and craters in the Martian polar caps. Icarus, Mars Polar Science VI 308, 138–147. https://doi.org/10.1016/j.icarus.2017.09.023

Ramirez, R.M., Kopparapu, R., Zugger, M.E., Robinson, T.D., Freedman, R., Kasting, J.F., 2014. Warming early Mars with CO2 and H2. Nat. Geosci. 7, 59–63. https://doi.org/10.1038/ngeo2000

Ramstad, R., Barabash, S., Futaana, Y., Nilsson, H., Holmström, M., 2018. Ion Escape From Mars Through Time: An Extrapolation of Atmospheric Loss Based on 10 Years of Mars Express Measurements. J. Geophys. Res. Planets 123, 3051–3060. https://doi.org/10.1029/2018JE005727

Ribas, I., Guinan, E.F., Güdel, M., Audard, M., 2005. Evolution of the Solar Activity over Time and Effects on Planetary Atmospheres. I. High-Energy Irradiances (1-1700 Å). Astrophys. J. 622, 680. https://doi.org/10.1086/427977

Sanz-Forcada, J., Micela, G., Ribas, I., Pollock, A.M.T., Eiroa, C., Velasco, A., Solano, E., García-Álvarez, D., 2011. Estimation of the XUV radiation onto close planets and their evaporation. Astron. Astrophys. 532, A6. https://doi.org/10.1051/0004-6361/201116594

Tanaka, K.L., Skinner, J.A., Jr., Dohm, J.M., Irwin, R.P., III, Kolb, E.J., Fortezzo, C.M., Platz, T., Michael, G.G., Hare, T.M., 2014. Geologic map of Mars. U.S. Geological Survey Scientific Investigations Map 3292.

Tu, L., Johnstone, C.P., Güdel, M., Lammer, H., 2015. The Extreme Ultraviolet and X-Ray Sun in Time: High-Energy Evolutionary Tracks of a Solar-Like Star. Astron. Astrophys. 577, L3. https://doi.org/10.1051/0004-6361/201526146

Urata, R.A., Toon, O.B., 2013. Simulations of the martian hydrologic cycle with a general circulation model: Implications for the ancient martian climate. Icarus 226, 229–250. https://doi.org/10.1016/j.icarus.2013.05.014

Usui, T., 2019. Chapter 4 - Hydrogen Reservoirs in Mars as Revealed by Martian Meteorites, in: Filiberto, J., Schwenzer, S.P. (Eds.), Volatiles in the Martian Crust. Elsevier, pp. 71–88. https://doi.org/10.1016/B978-0-12-804191-8.00004-0

Villanueva, G.L., Mumma, M.J., Novak, R.E., Käufl, H.U., Hartogh, P., Encrenaz, T., Tokunaga, A., Khayat, A., Smith, M.D., 2015. Strong water isotopic anomalies in the martian atmosphere: Probing current and ancient reservoirs. Science 348, 218–221. https://doi.org/10.1126/science.aaa3630

Vos, E., Aharonson, O., Schorghofer, N., 2019. Dynamic and isotopic evolution of ice reservoirs on Mars. Icarus 324, 1–7.

Warner, N.H., Golombek, M.P., Sweeney, J., Fergason, R., Kirk, R., Schwartz, C., 2017. Near Surface Stratigraphy and Regolith Production in Southwestern Elysium Planitia, Mars: Implications for Hesperian-Amazonian Terrains and the InSight Lander Mission. Space Sci. Rev. 211, 147–190. https://doi.org/10.1007/s11214-017-0352-x

Warren, A.O., Kite, E.S., Williams, J.-P., Horgan, B., n.d. Through the Thick and Thin: New Constraints on Mars Paleopressure History 3.8 – 4 Ga from Small Exhumed Craters. J. Geophys. Res. Planets n/a. https://doi.org/10.1029/2019JE006178

Webster, C.R., Mahaffy, P.R., Flesch, G.J., Niles, P.B., Jones, J.H., Leshin, L.A., Atreya, S.K., Stern, J.C., Christensen, L.E., Owen, T., Franz, H., Pepin, R.O., Steele, A., Team, the M.S., 2013. Isotope Ratios of H, C, and O in CO2 and H2O of the Martian Atmosphere. Science 341, 260–263. https://doi.org/10.1126/science.1237961

Wernicke, L.J., Jakosky, B.M., 2019. The History of Water on Mars: Hydrated Minerals as a Water Sink in the Martian Crust. LPI Contrib. 2089.

Whitten, J.L., Campbell, B.A., Plaut, J.J., 2018. Defining the material properties of the Dorsa Argentea Formation using MARSIS radar sounder data. LPSC XLIX #2632.

Wordsworth, R.D., 2016. The Climate of Early Mars. Annu. Rev. Earth Planet. Sci. 44, 381–408. https://doi.org/10.1146/annurev-earth-060115-012355

| Table 1: Ages of Martian epoch boundaries | |
|---|---|
| Epoch | Start of epoch (Ga) |
| Late Amazonian | 0.274 |
| Middle Amazonian | 1.03 |
| Early Amazonian | 3.24 |
| Late Hesperian | 3.39 |
| Early Hesperian | 3.56 |
| Late Noachian | 3.85 |
| Middle Noachian | 3.96 |
| Early Noachian | N/A |
| Chronology data after (Michael, 2013), making use of the 'Hartmann 2004' iteration in that study. | |

| Table 2: List of input parameters for Mars 3.5 Gyr volatile evolution model described by Equation 1 | | | |
|---|---|---|---|
| Parameter | Symbol | Value | Distribution in range |
| Modern dissociative recombination O loss (atoms/sec) | $R_{present}$ | $4.3^{+5.2}_{-2.3} \times 10^{25}$ (1σ)[1] | Log-normal |
| Solar EUV evolution exponent | $\beta_{rot}$ | $1.22^{+0.96}_{-0.26}$ (10th, 90th percentiles)[2] | Half normal about mean |
| | $\beta_{lum}$ | $1.24 \pm 0.14$ (1σ)[3] | Normal |
| EUV flux dependence exponent | $\gamma_{2.64}$ | $2.64 \pm 0.60$ (1σ)[1] | Normal |
| | $\gamma_{1.7}$ | $1.70 \pm 0.39$ (1σ)[4] | Normal |
| | $\gamma_{1.0}$ | $1.0 \pm 0.23$ (1σ)[5] | Normal |
| Martian 3.5 Ga D/H, *in situ* mudstone | $I_{t1}$ | $3.0 \pm 0.2 \times$ SMOW (2σ)[6] | Normal |
| Martian modern D/H, *in situ* | $I_{t2}$ | $6.0 \pm 1 \times$ SMOW (2σ)[7] | Normal |
| Martian modern D/H, global | | $8.0 \pm 1 \times$ SMOW (2σ)[8*] | Normal |
| Fractionation factor | F | 0.016-0.33[9] | Uniform |
| Martian modern H$_2$O GEL (m) | $R_{t2}$ | $31.5 \pm 1.25$ (2σ)[10] | Normal |
| Martian modern exchangeable CO$_2$ (mbar) | $pCO_{2(modern)}$ | 12[11] | N/A |
| Martian post 3.5 Ga soil and sediment O sink | $O_{soil}$ | 242 mbar pCO$_2$ equivalent/5.34 m GEL H$_2$O[12] | N/A |
| 3.5 Ga total CO$_2^+$ loss | #CO$_2^+_{esc}$ | 2.1 mbar pCO$_2$ equivalent[13] | N/A |

[1] Lillis et al. (2017)
[2] Tu et al. (2015), stellar rotational modeling, uncertainties as two half-normal distributions about mean
[3] Sanz-Forcada et al. (2011), fit to X-ray luminosity observations
[4] R. J. Lillis, personal communication. Fractional error in exponent assumed same as 1
[5] Cravens et al. (2018). Proportional error assumed same as [1]
[6] Mahaffy et al. (2015)
[7] Webster et al. (2013)
[8] Villanueva et al. (2015), error bars assumed same as 6
[9] Kurokawa et al. (2016), uniform distribution within range assumed
[10] Carr and Head (2015), error bars assigned based on upper and lower limits described
[11] Putzig et al. (2018), accounts for seasonal CO$_2$ ice deposits and atmospheric reservoir
[12] Held fixed at estimated value from Supplementary Table 1
[13] Dong et al. (2018)

| Table 3: Simulated 3.5 Ga paleo-pCO$_2$ for different input parameters | | | | | | | |
|---|---|---|---|---|---|---|---|
| Modern D/H = 6.0 ± 1 x SMOW | | | | | | | |
| | 3.5 Ga paleo pCO$_2$ (mbar) | | | | | | |
| | Significance level: full distribution | | | Significance level: discarding <12 mbar pCO$_2$ outcomes | | | % >12 mbar |
| EUV model | 0.05 | 0.50 | 0.95 | 0.05 | 0.50 | 0.95 | |
| $\gamma_{2.64}\ \beta_{rot}$ | -1846 | -369 | 83442 | 153 | 4137 | 363325 | 42.2 |
| $\gamma_{1.7}\ \beta_{rot}$ | -2132 | -1032 | 2501 | 89 | 1430 | 25169 | 13.5 |
| $\gamma_{1.0}\ \beta_{rot}$ | -2281 | -1254 | -460 | 43 | 499 | 4267 | 1.4 |
| $\gamma_{2.64}\ \beta_{lum}$ | -1884 | -614 | 4835 | 88 | 1296 | 11153 | 30.6 |
| $\gamma_{1.7}\ \beta_{lum}$ | -2181 | -1134 | -124 | 46 | 450 | 2741 | 3.7 |
| $\gamma_{1.0}\ \beta_{lum}$ | -2305 | -1288 | -583 | 30 | 226 | 1300 | 0.03 |
| | | | | | | | |
| Modern D/H = 8.0 ± 1 x SMOW | | | | | | | |
| $\gamma_{2.64}\ \beta_{rot}$ | -3714 | -1639 | 81953 | 310 | 8088 | 626049 | 29.9 |
| $\gamma_{1.7}\ \beta_{rot}$ | -4077 | -2347 | 1158 | 153 | 2932 | 44652 | 7.0 |
| $\gamma_{1.0}\ \beta_{rot}$ | -4261 | -2620 | -1567 | 65 | 967 | 7183 | 0.3 |
| $\gamma_{2.64}\ \beta_{lum}$ | -3778 | -1910 | 3458 | 131 | 1963 | 16165 | 15.0 |
| $\gamma_{1.7}\ \beta_{lum}$ | -4142 | -2487 | -1289 | 70 | 683 | 4143 | 0.7 |
| $\gamma_{1.0}\ \beta_{lum}$ | -4281 | -2659 | -1668 | 59 | 757 | 2275 | 0.001 |

**Figures**

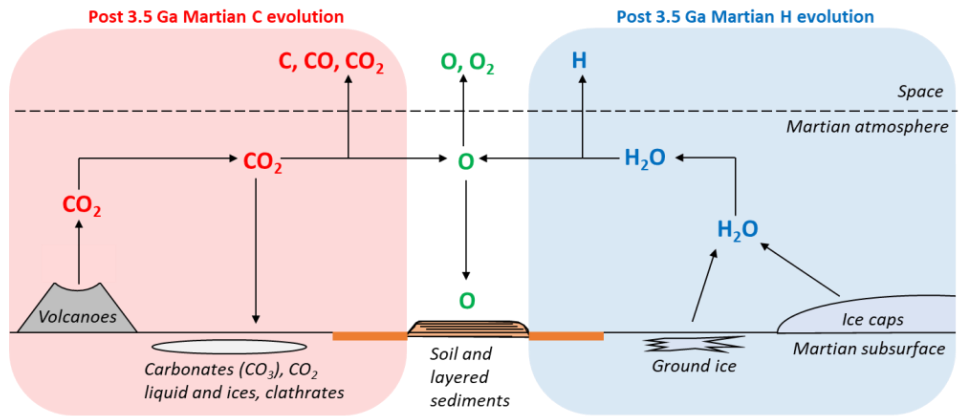

1A

$$\#C_{esc,max} = 1/2 \times \left( \#O_{esc,max} - \left( 1/2 \times \#H_{esc,min} - \#O_{ox,max} \right) \right) + \#CO_2^+{}_{esc,max}$$

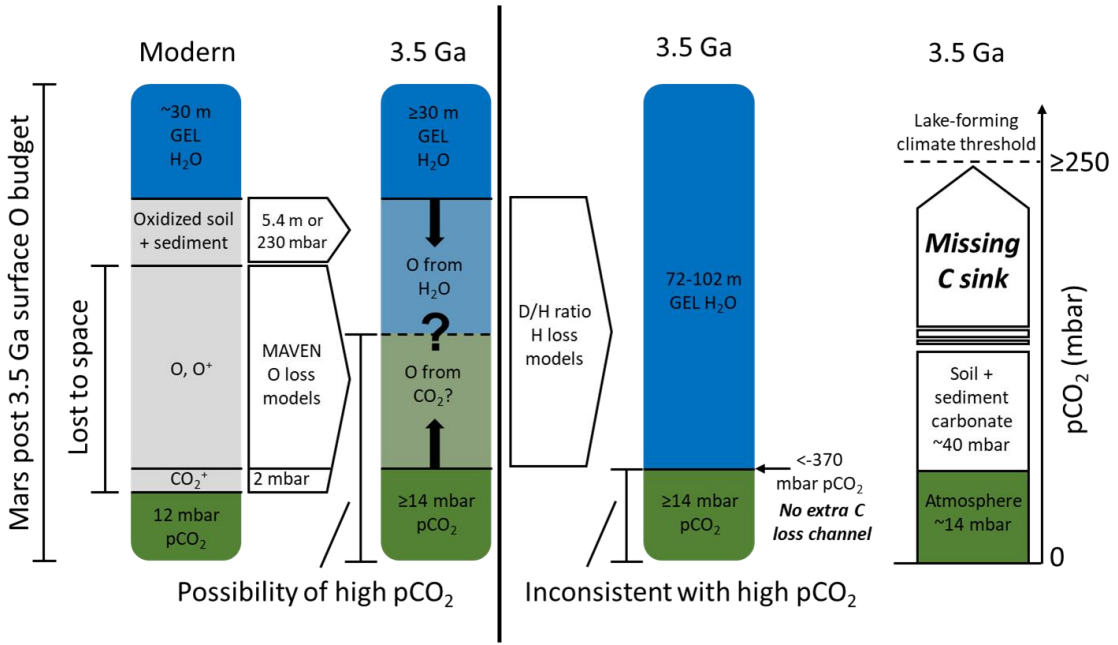

1B

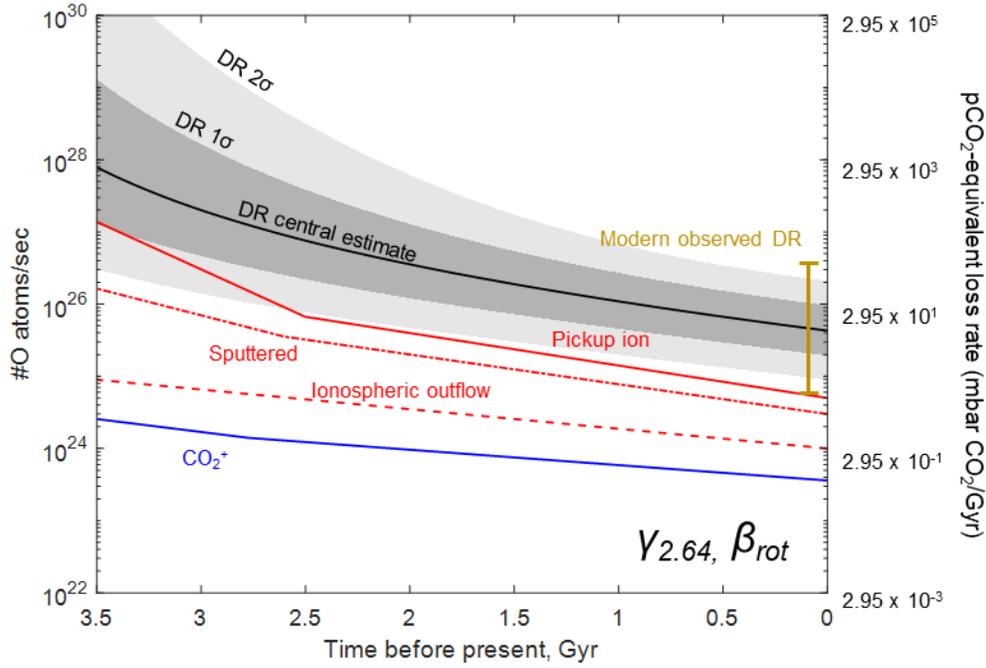

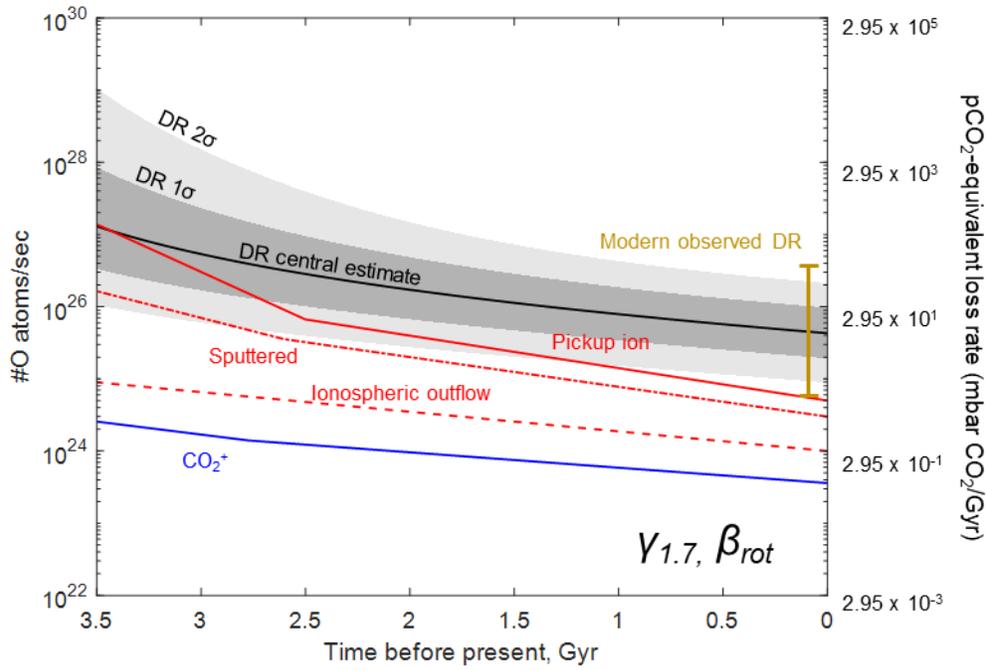

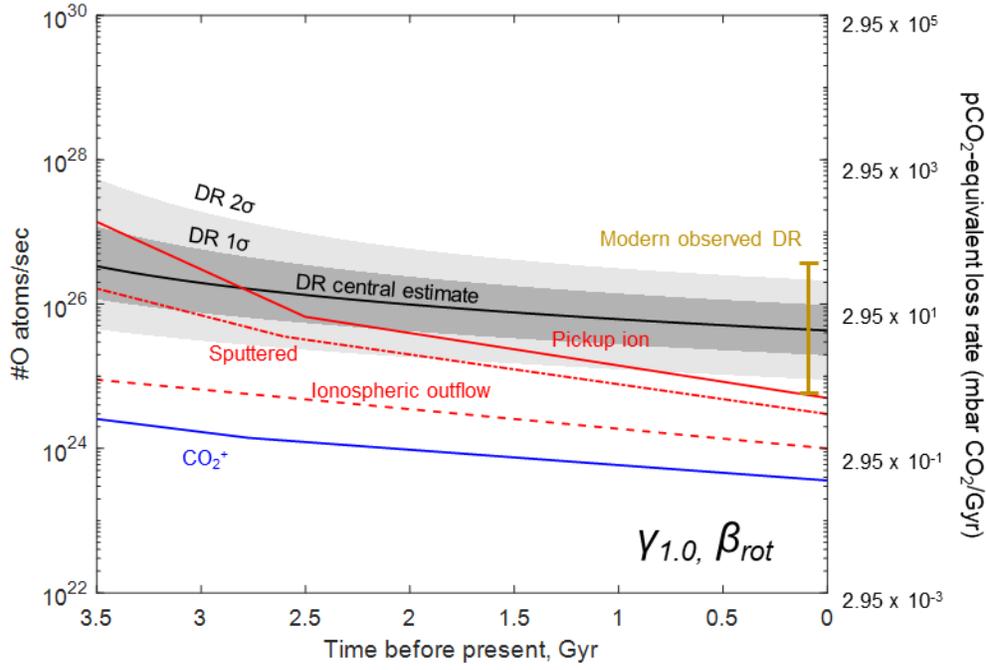

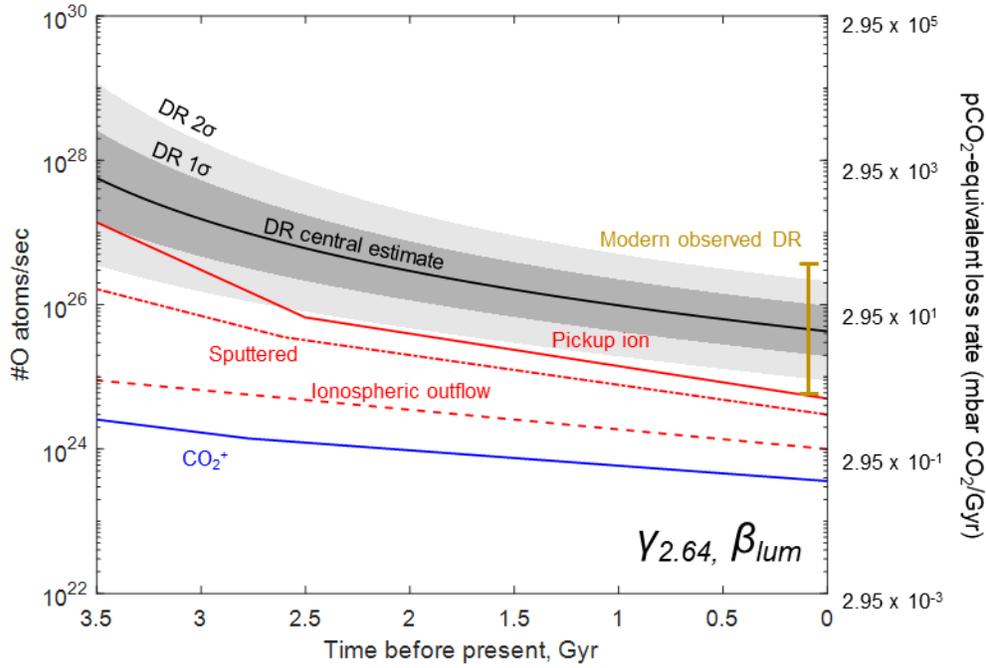

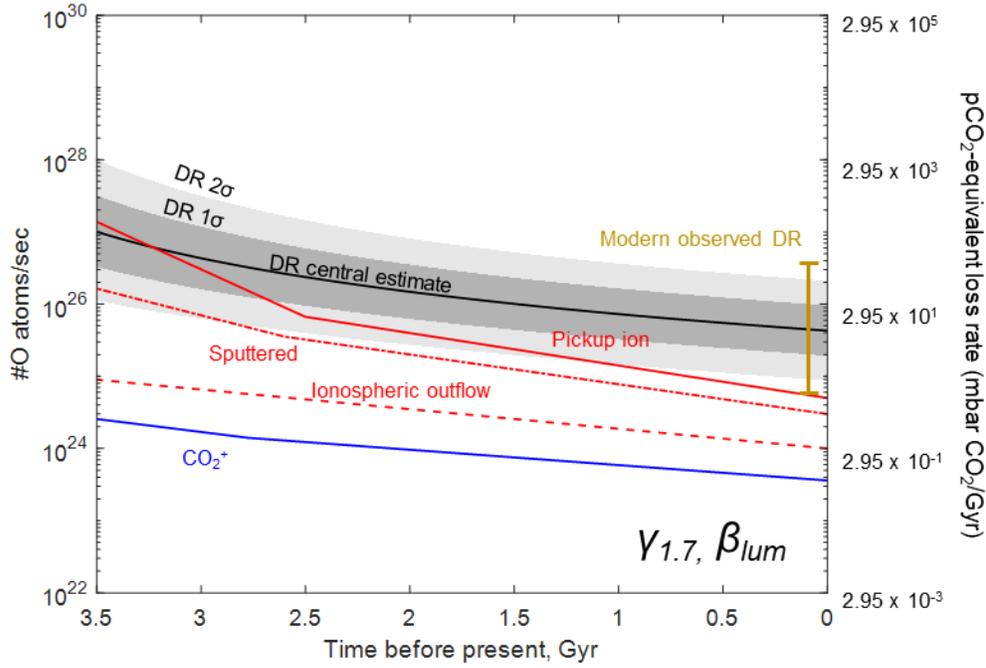
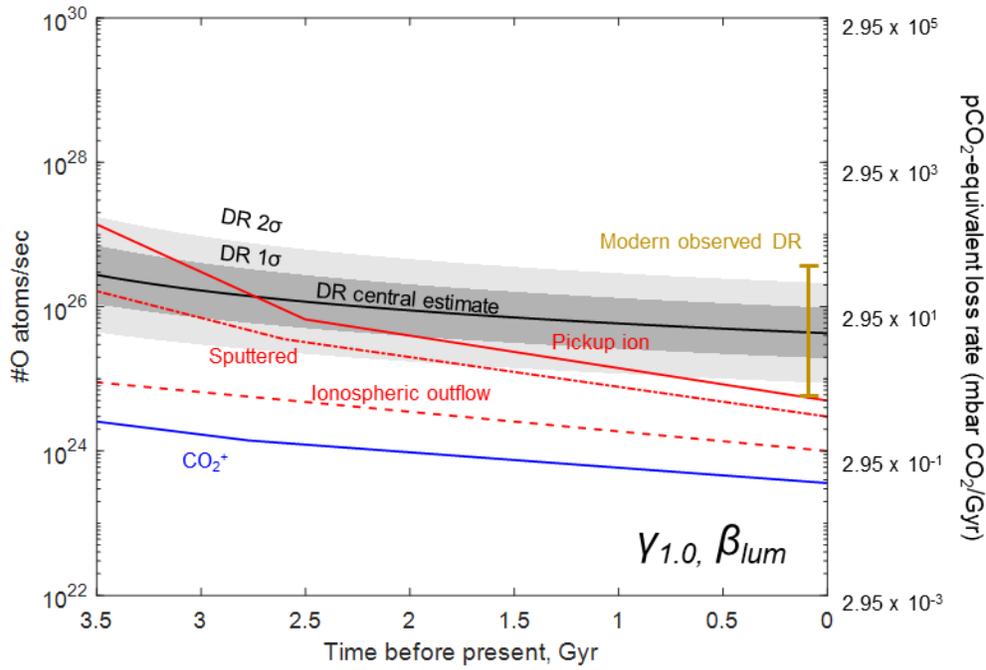

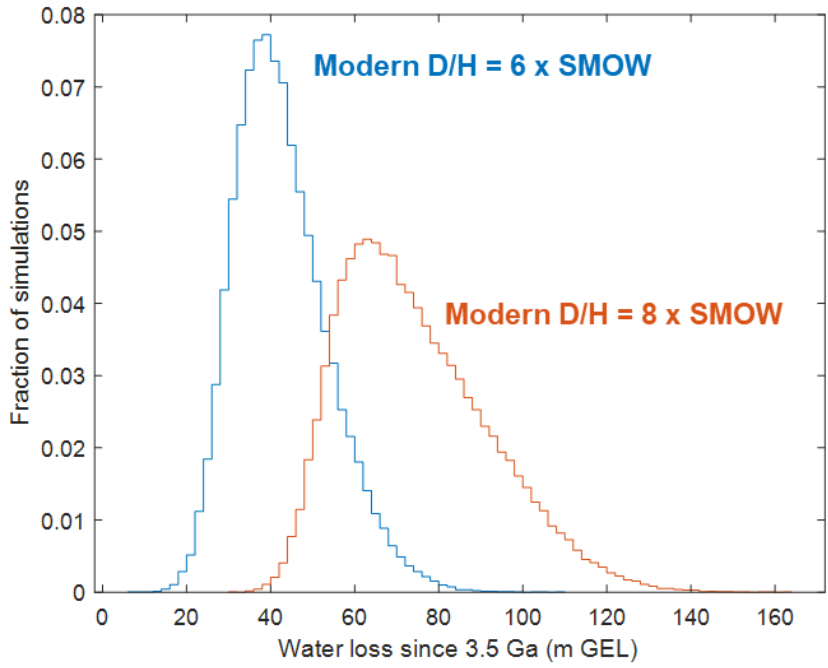

4A

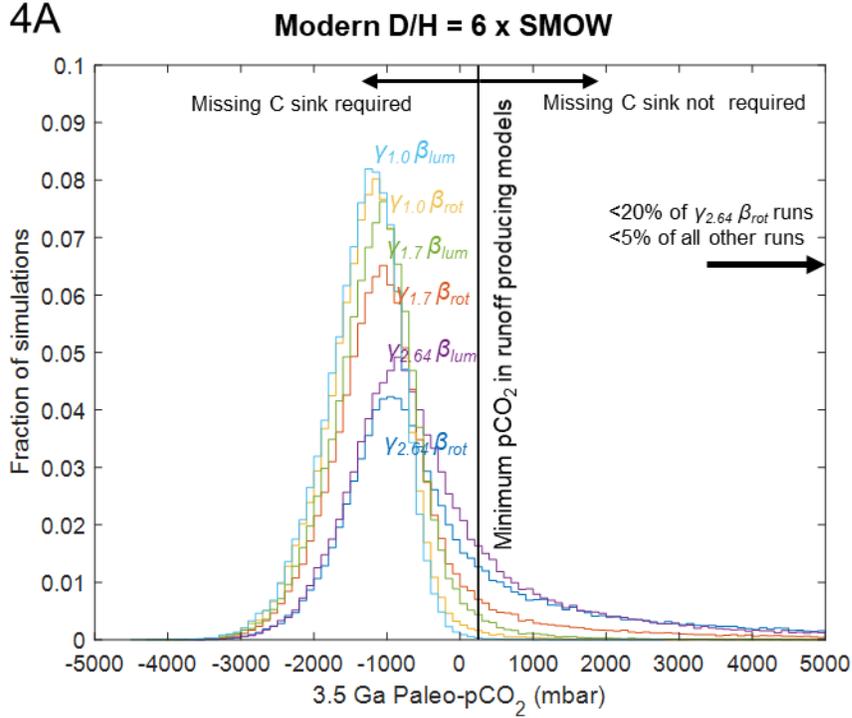

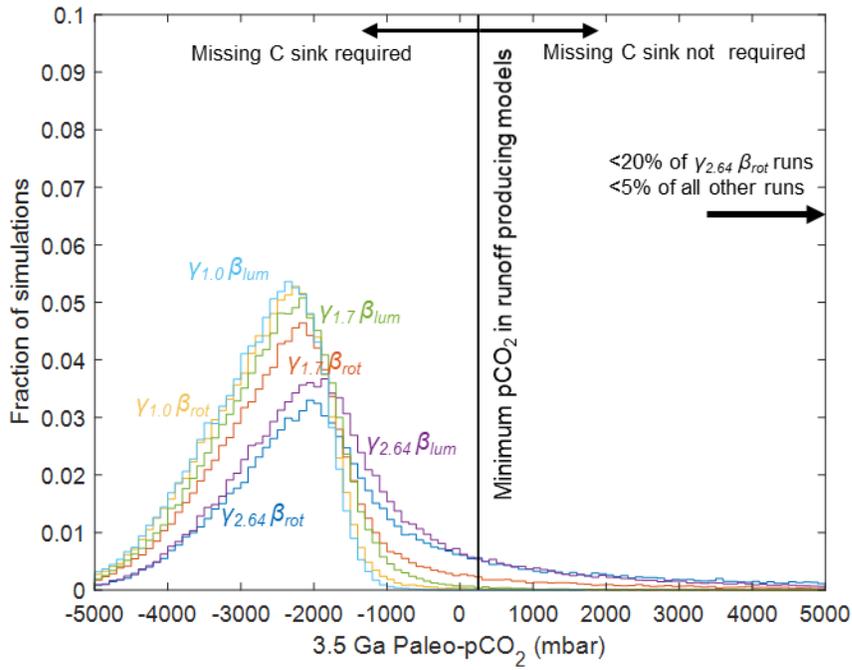
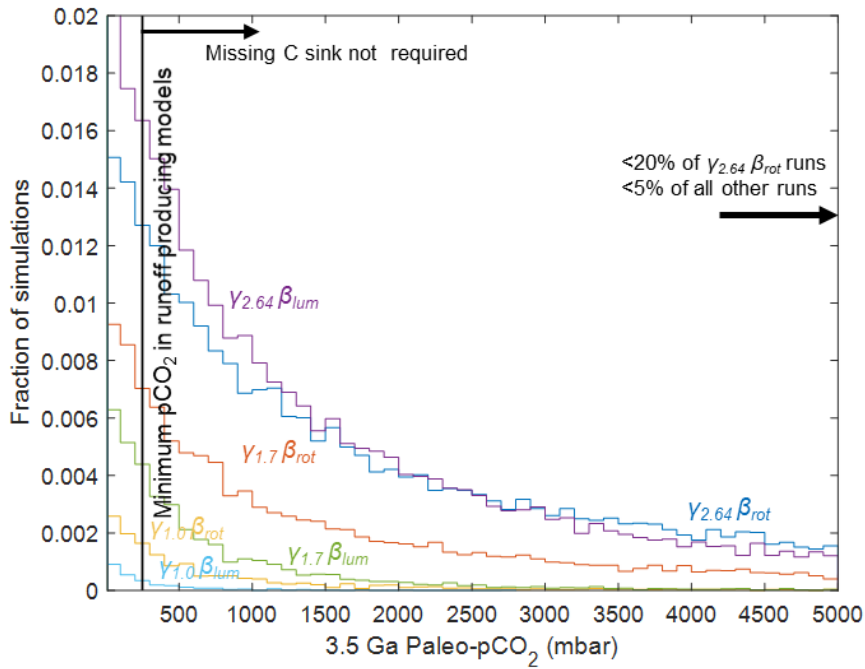

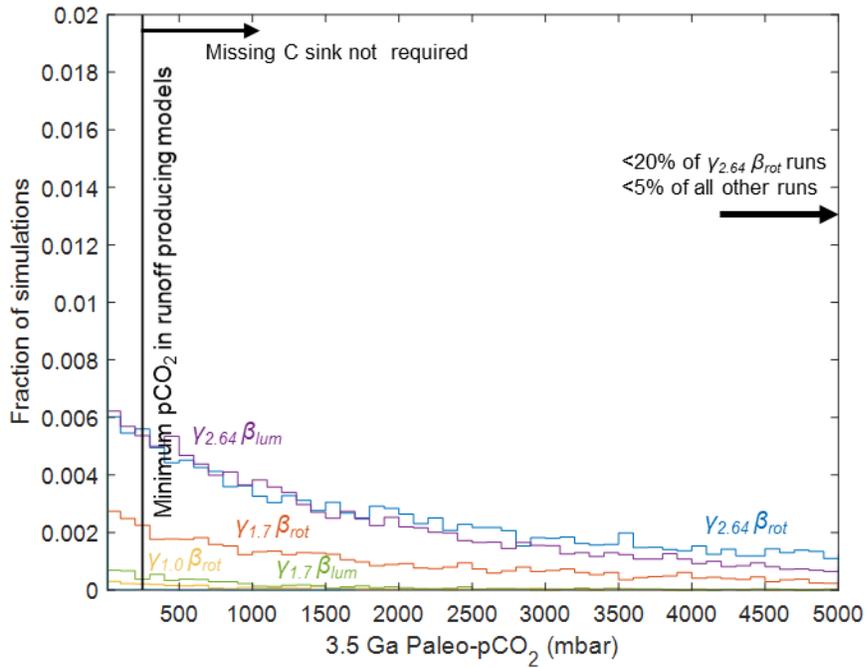

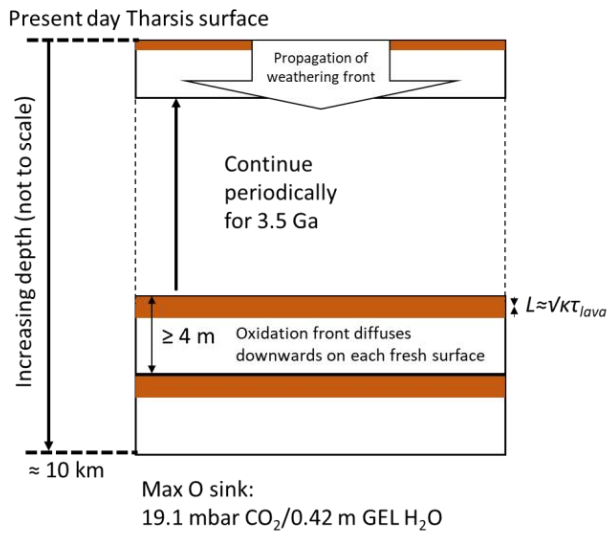

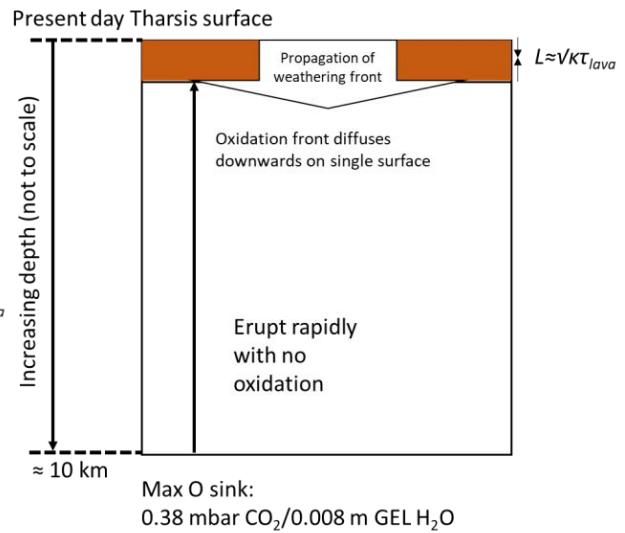

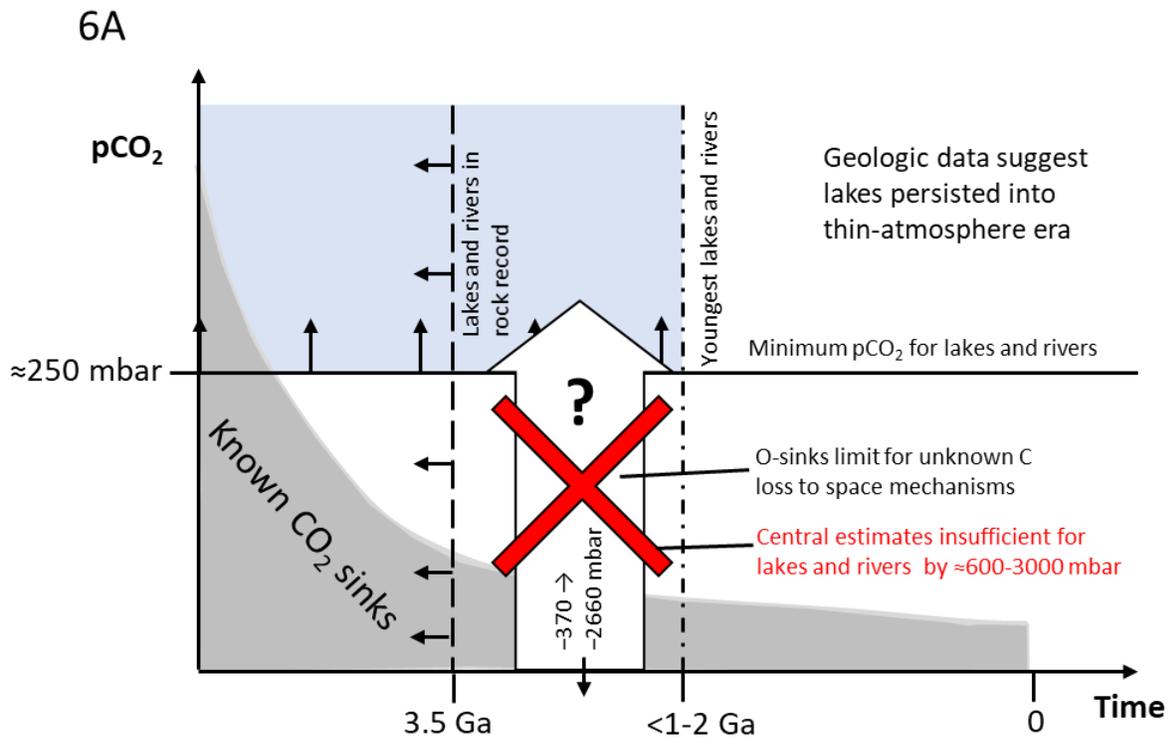
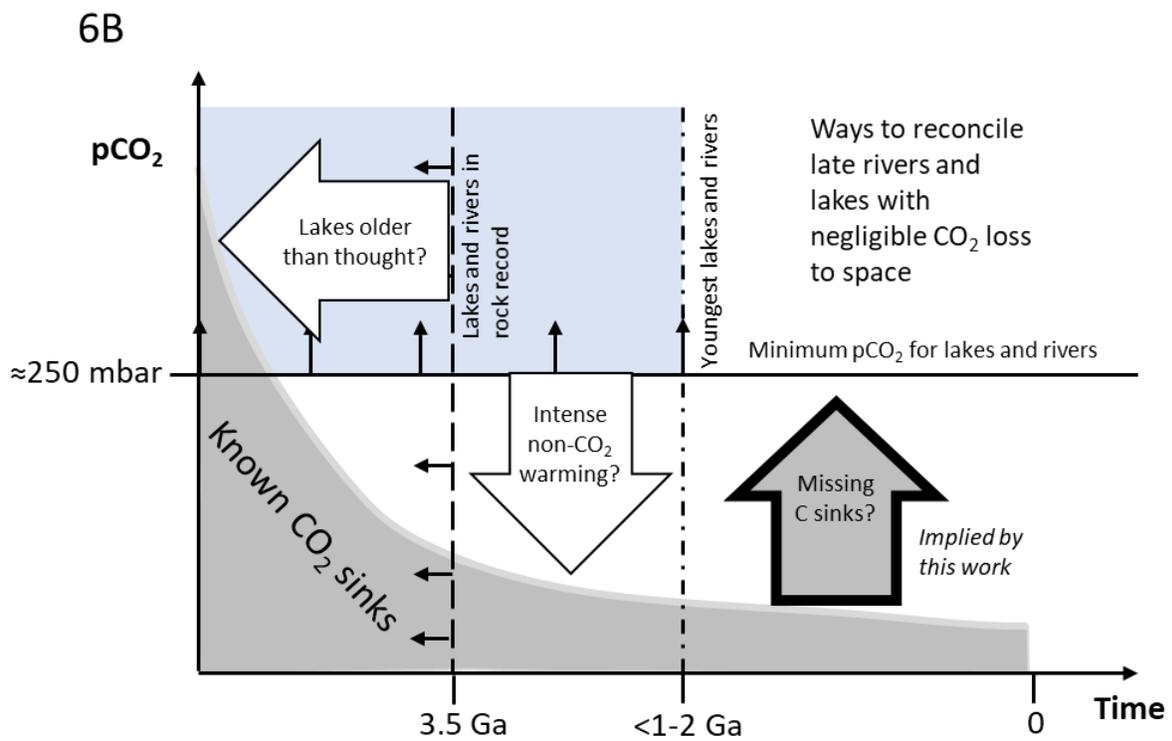

| Supplementary Table 1: Young sedimentary reservoirs on Mars and their estimated O sinks | | | | | | |
|---|---|---|---|---|---|---|
| Unit | Volume (m$^3$) | Density (kg/m$^3$) | % Dust[1] | Mass dust (kg) | Added O (kg) | Reference |
| Dunes | 9.0E+12 | 1890 | 100 | 1.7E+16 | 1.54E+15 | Hayward (2011); Hayward et al. (2012, 2007) |
| SPLD | 1.6E+15 | 1220 | 15 | 2.9E+17 | 2.65E+16 | Plaut et al. (2007); Zuber et al. (2007) |
| Duststone | 1.2E+14[2] | 1500 | 100 | 1.8E+17 | 1.62E+16 | Bridges et al. (2010) |
| Medusae Fossae Formation | 1.9E+15 | 1800 | 100 | 3.4E+18 | 3.10E+17 | Watters et al. (2007) |
| North polar layered deposits, excluding basal unit | 8.2E+14 | 1000 | 5 | 4.1E+16 | 3.72E+15 | Grima et al. (2009) |
| Basal Unit, NPLD | 2.7E+14 | 1960 | 77 | 4.0E+17 | 3.67E+16 | Byrne and Murray (2002) |
| Mid latitude mantling deposits | 1.3E+15 | 1960 | 77 | 1.9E+18 | 1.77E+17 | Byrne et al. (2009); Schon et al. (2009) |
| Low latitude glacial ice | 6.0E+14 | 1000 | 10 | 6.0E+16 | 5.43E+15 | Shean et al. (2005) |
| Near-surface ejecta | 3.6E+12 | 2000 | 100 | 7.2E+15 | 6.52E+14 | Scott and Tanaka (1986) |
| Global soil and dust | 4.4E+14[3] | 1500 | 100 | 6.5E+16 | 5.91E+15 | Warner et al. (2017) |
| Southern mid latitude icy deposits | 2.8E+13 | 1000 | 10 | 2.8E+15 | 2.54E+14 | Holt et al. (2008) |
| Interior layered deposits | 2.4E+14 | 2400 | 100 | 5.8E+17 | 5.22E+16 | Michalski and Niles (2012) |
| Total O mass (kg) | | | | | 6.89E+17 | |
| Total O moles | | | | | 4.30E+19 | |

Added O was calculated assuming modern $Fe^{3+}/Fe_{Total}=0.39$, and 20 wt% $SO_3$ derived from 50% $H_2S$, 50% $SO_2$

[1] Maximum dust content was estimated for formations with <100% dust by the authors, assuming an admixture of silicate-dominated soil material with pure water ice, except in the case of the SPLD where the dust content is taken from the published estimate of Zuber et al. (2007). Only the soil component of dust/ice mixtures represents an oxidative sink for O.

[2] Maximum, assuming 4 m thickness on 20.5% of Mars surface

[3] Assuming regolith thickness of 3 m

# References

Bridges, N.T., Banks, M.E., Beyer, R.A., Chuang, F.C., Noe Dobrea, E.Z., Herkenhoff, K.E., Keszthelyi, L.P., Fishbaugh, K.E., McEwen, A.S., Michaels, T.I., Thomson, B.J., Wray, J.J., 2010. Aeolian bedforms, yardangs, and indurated surfaces in the Tharsis Montes as seen by the HiRISE Camera: Evidence for dust aggregates. Icarus, MRO/HiRISE Studies of Mars 205, 165–182. https://doi.org/10.1016/j.icarus.2009.05.017

Byrne, S., Dundas, C.M., Kennedy, M.R., Mellon, M.T., McEwen, A.S., Cull, S.C., Daubar, I.J., Shean, D.E., Seelos, K.D., Murchie, S.L., Cantor, B.A., Arvidson, R.E., Edgett, K.S., Reufer, A., Thomas, N., Harrison, T.N., Posiolova, L.V., Seelos, F.P., 2009. Distribution of Mid-Latitude Ground Ice on Mars from New Impact Craters. Science 325, 1674–1676. https://doi.org/10.1126/science.1175307

Byrne, S., Murray, B.C., 2002. North polar stratigraphy and the paleo-erg of Mars. J. Geophys. Res. Planets 107, 11-1-11–12. https://doi.org/10.1029/2001JE001615

Grima, C., Kofman, W., Mouginot, J., Phillips, R.J., Hérique, A., Biccari, D., Seu, R., Cutigni, M., 2009. North polar deposits of Mars: Extreme purity of the water ice. Geophys. Res. Lett. 36. https://doi.org/10.1029/2008GL036326

Hayward, R.K., 2011. Mars Global Digital Dune Database (MGD3): north polar region (MC-1) distribution, applications, and volume estimates. Earth Surf. Process. Landf. 36, 1967–1972. https://doi.org/10.1002/esp.2219

Hayward, R.K., Fenton, L.K., Titus, T.N., Colaprete, A., Christensen, P.R., 2012. SP_Dune_Pamphlet.doc. Mars Glob. Digit. Dune Database.

Hayward, R.K., Mullins, K.F., Fenton, L.K., Hare, T.M., Titus, T.N., Bourke, M.C., Colaprete, A., Christensen, P.R., 2007. Mars Global Digital Dune Database and initial science results. J. Geophys. Res. Planets 112. https://doi.org/10.1029/2007JE002943

Holt, J.W., Safaeinili, A., Plaut, J.J., Head, J.W., Phillips, R.J., Seu, R., Kempf, S.D., Choudhary, P., Young, D.A., Putzig, N.E., Biccari, D., Gim, Y., 2008. Radar Sounding Evidence for Buried Glaciers in the Southern Mid-Latitudes of Mars. Science 322, 1235–1238. https://doi.org/10.1126/science.1164246

Michalski, J., Niles, P.B., 2012. Atmospheric origin of Martian interior layered deposits: Links to climate change and the global sulfur cycle. Geology 40, 419–422. https://doi.org/10.1130/G32971.1

Plaut, J.J., Picardi, G., Safaeinili, A., Ivanov, A.B., Milkovich, S.M., Cicchetti, A., Kofman, W., Mouginot, J., Farrell, W.M., Phillips, R.J., Clifford, S.M., Frigeri, A., Orosei, R., Federico, C., Williams, I.P., Gurnett, D.A., Nielsen, E., Hagfors, T., Heggy, E., Stofan, E.R., Plettemeier, D., Watters, T.R., Leuschen, C.J., Edenhofer, P., 2007. Subsurface Radar Sounding of the South Polar Layered Deposits of Mars. Science 316, 92–95. https://doi.org/10.1126/science.1139672

Schon, S.C., Head, J.W., Milliken, R.E., 2009. A recent ice age on Mars: Evidence for climate oscillations from regional layering in mid-latitude mantling deposits. Geophys. Res. Lett. 36. https://doi.org/10.1029/2009GL038554

Scott, D.H., Tanaka, K.L., 1986. Geologic map of the western equatorial region of Mars. IMAP. https://doi.org/10.3133/i1802A

Shean, D.E., Head, J.W., Marchant, D.R., 2005. Origin and evolution of a cold-based tropical mountain glacier on Mars: The Pavonis Mons fan-shaped deposit. J. Geophys. Res. Planets 110. https://doi.org/10.1029/2004JE002360

Warner, N.H., Golombek, M.P., Sweeney, J., Fergason, R., Kirk, R., Schwartz, C., 2017. Near Surface Stratigraphy and Regolith Production in Southwestern Elysium Planitia, Mars: Implications for Hesperian-Amazonian Terrains and the InSight Lander Mission. Space Sci. Rev. 211, 147–190. https://doi.org/10.1007/s11214-017-0352-x

Watters, T.R., Campbell, B., Carter, L., Leuschen, C.J., Plaut, J.J., Picardi, G., Orosei, R., Safaeinili, A., Clifford, S.M., Farrell, W.M., Ivanov, A.B., Phillips, R.J., Stofan, E.R., 2007. Radar Sounding of the Medusae Fossae Formation Mars: Equatorial Ice or Dry, Low-Density Deposits? Science 318, 1125–1128. https://doi.org/10.1126/science.1148112

Zuber, M.T., Phillips, R.J., Andrews-Hanna, J.C., Asmar, S.W., Konopliv, A.S., Lemoine, F.G., Plaut, J.J., Smith, D.E., Smrekar, S.E., 2007. Density of Mars' South Polar Layered Deposits. Science 317, 1718–1719. https://doi.org/10.1126/science.1146995